\documentclass[twocolumn]{aastex63}

\usepackage{textcomp}
\usepackage{comment}
\usepackage{lineno}

\accepted{ApJ}

\shorttitle{Double-peaked emitters from the Zwicky Transient Facility}
\shortauthors{Ward et al.}

\graphicspath{{./}{figures/}}

\begin{document}

\title{Panic at the ISCO: time-varying double-peaked broad lines from evolving accretion disks are common amongst optically variable AGN}

\correspondingauthor{Charlotte Ward}
\email{charlotte.ward@princeton.edu}

\author[0000-0002-4557-6682]{Charlotte Ward}
\affil{Department of Astrophysical Sciences, Princeton University, Princeton, NJ 08544, USA} 
\author[0000-0002-0786-7307]{Suvi Gezari}
\affil{Space Telescope Science Institute, 3700 San Martin Dr., Baltimore, MD 21218, USA}
\author[0000-0002-3389-0586]{Peter Nugent}
\affiliation{Lawrence Berkeley National Laboratory, 1 Cyclotron Road, Berkeley, CA 94720, USA}
\affiliation{Department of Astronomy, University of California, Berkeley, Berkeley, CA 94720, USA}
\author[0000-0002-0893-4073]{Matthew Kerr}
\affiliation{Space Science Division, U.S. Naval Research Laboratory, Washington, DC 20375, USA}
\author[0000-0002-3719-940X]{Michael Eracleous}
\affiliation{Department of Astronomy \& Astrophysics and Institute for Gravitation and the Cosmos, 525 Davey Laboratory, The Pennsylvania State University, University Park, PA 16802, USA}
\author[0000-0001-9676-730X]{Sara Frederick}
\affiliation{Department of Physics and Astronomy, Vanderbilt University, 6301 Stevenson Center, Nashville, TN 37235}
\author[0000-0002-5698-8703]{Erica Hammerstein}
\affil{Department of Astronomy, University of Maryland, College Park, MD  20742, USA}
\author[0000-0002-3168-0139]{Matthew J. Graham}
\affiliation{Division of Physics, Mathematics, and Astronomy, California Institute of Technology, Pasadena, CA 91125, USA}
\author[0000-0002-3859-8074]{Sjoert van Velzen}
\affil{Leiden Observatory, Leiden University, Postbus 9513, 2300 RA, Leiden, the Netherlands}
\author[0000-0002-5619-4938]{Mansi M. Kasliwal}
\affil{Division of Physics, Mathematics, and Astronomy, California Institute of Technology, Pasadena, CA 91125, USA}
\author[0000-0003-2451-5482]{Russ R. Laher}
\affiliation{IPAC, California Institute of Technology, 1200 E. California
             Blvd, Pasadena, CA 91125, USA}
\author[0000-0002-8532-9395]{Frank J. Masci}
\affiliation{IPAC, California Institute of Technology, 1200 E. California
             Blvd, Pasadena, CA 91125, USA}
\author[0000-0003-1227-3738]{Josiah Purdum}
\affil{Caltech Optical Observatories, California Institute of Technology, Pasadena, CA 91125, USA}
\author[0000-0003-1227-3738]{Benjamin Racine}
\affil{Aix Marseille Univ, CNRS/IN2P3, CPPM, Marseille, France}
\author[0000-0001-7062-9726]{Roger Smith}
\affil{Caltech Optical Observatories, California Institute of Technology, Pasadena, CA  91125}

\begin{abstract}
About 3-10\% of Type I active galactic nuclei (AGN) have double-peaked broad Balmer lines in their optical spectra originating from the motion of gas in their accretion disk. Double-peaked profiles arise not only in AGN, but occasionally appear during optical flares from tidal disruption events and changing-state AGN. In this paper we identify 250 double-peaked emitters (DPEs) amongst a parent sample of optically variable broad-line AGN in the Zwicky Transient Facility (ZTF) survey, corresponding to a DPE fraction of 19\%. We model spectra of the broad H$\alpha$ emission line regions and provide a catalog of the fitted accretion disk properties for the 250 DPEs. Analysis of power spectra derived from the 5 year ZTF light curves finds that DPE light curves have similar amplitudes and power law indices to other broad-line AGN. Follow-up spectroscopy of 12 DPEs reveals that $\sim$50\% display significant changes in the relative strengths of their red and blue peaks over long $10-20$ year timescales, indicating that broad-line profile changes arising from spiral arm or hotspot rotation are common amongst optically variable DPEs. Analysis of the accretion disk parameters derived from spectroscopic modeling provides evidence that DPEs are not in a special accretion state, but are simply normal broad-line AGN viewed under the right conditions for the accretion disk to be easily visible. We include inspiraling SMBH binary candidate SDSSJ1430+2303 in our analysis, and discuss how its photometric and spectroscopic variability is consistent with the disk-emitting AGN population in ZTF.

\vspace{1cm}
\end{abstract}

\section{Introduction}
The vast majority of massive galaxies host a supermassive black hole (SMBH) in their center \citep{Magorrian1998TheCenters}. Understanding how efficiently SMBHs grow via accretion of gas in galaxy nuclei is essential if we are to determine how SMBHs formed in the early Universe and how they have co-evolved with their host galaxies over time \citep{Pacucci2015TheHoles}. Direct emission from the gas disks around SMBHs has been observed in optical spectra of some galactic nuclei and provides important observational data for comparison to simulations of SMBH accretion and for understanding the efficiency of various active galactic nuclei (AGN) accretion states. 

Emission from AGN accretion disks is sometimes observable as broad double-peaked H$\alpha$ and H$\beta$ emission lines with each peak at $\pm (500-3000)$ km/s from the rest velocity \citep{Eracleous1994Doubled-peakedNuclei,Chen1989, Eracleous1997RejectionNucleib,Strateva2003Double-peakedNuclei,Eracleous2009Double-peakedNuclei}. AGN with double-peaked broad Balmer emission lines are called double-peaked emitters (DPEs). The double-peaked Balmer lines are usually well-modeled as emission from a geometrically thin and optically thick relativistic Keplerian accretion disk, where Doppler boosting results in asymmetry between the red and blue peaks \citep{Chen1989,Strateva2003Double-peakedNuclei}. Factors such as turbulent broadening and the emissivity profile of the disk generate a variety of disk profile shapes. As the emitting region of the disk producing the double-peaked H$\alpha$ and H$\beta$ profiles is of the order of tens to hundreds of gravitational radii for known DPEs, disk models applied to double-peaked Balmer profiles cannot probe the innermost stable orbit (ISCO), but instead provide information about the outer regions of the accretion disk from tens to thousands of graviational radii.

Over the last three decades, increasing numbers of AGN have been found to exhibit long-lived double-peaked disk emission from a stable accretion disk \citep[e.g. the canonical DPE Arp102B;][]{Chen1989Kinematic102B, Popovic2014}. Estimates of DPE fractions amongst the wider broad-line AGN population range from $\sim 3-30$\% \citep{Eracleous1994Doubled-peakedNuclei,Ho1997AEmission,Strateva2003Double-peakedNuclei}.  Some DPEs show substantial changes in the relative flux of the blue and red peaks over timescales of years to decades which is well modeled by the rotation of spiral arms or hotspots in the disk \citep{Storchi-Bergmann2002Double-peaked1097,Lewis2010Long-termLines,Gezari2007LongTermNuclei,Schimoia2012ShortNGC1097,Schimoia2017EvolutionNGC7213}. 

Over the past few years, spectroscopic follow-up of transient phenomena in wide-field optical time-domain surveys has revealed new classes of disk-emitters: those with transient double-peaked emission from a temporary accretion disk associated with tidal disruption of a star or the onset of a new AGN accretion episode in a previously inactive AGN. Tidal disruption events (TDEs) occur when a star passing an SMBH is disrupted by tidal forces, causing the formation a disk of material which is partially accreted onto the SMBH and produces a flare in the optical, UV and X-rays \citep{Rees1988TidalGalaxies,Evans1989TheHole,Ulmer1999FlaresHoles}. Some recently reported TDEs have exhibited the appearance of a double-peaked Balmer profile associated with the onset of the initial optical flare, which in some cases has been followed by fading of the disk profile over the following year \citep{Short2020TheDecrement, Nicholl2020AnAT2019qiz, Hung2020Double-peakedEvent,Holoien2019PS18kh:Disk}. The periodic nuclear transient ASASSN-14ko, likely a repeating partial TDE \citep{Payne2021ASASSN-14ko253-G003}, exhibited a double-peaked spectrum consistent with a circular disk containing a spiral arm, and the calculated precession timescale of that spiral arm matched the periodic flaring timescale of $\sim 114$ days \citep{Tucker2021AnAGN}. 

Double-peaked and asymmetric broad lines have also been observed amongst populations of changing-state AGN (or changing-look AGN; CLAGN). Changing-state AGN are identified by the appearance or disappearance of broad Balmer emission lines indicating a change in the presence of gas in the vicinity of the SMBH, and are often associated with the onset or termination of optical variability \citep[see][for a review of changing-state AGN]{Ricci2023Changing-lookNuclei}. Changing-look LINERS NGC 1097 and NGC 3065 have irregularly shaped broad-line profiles which may be ascribed to accretion disk emission \citep{StorchiBergmann2003EvolutionHole,Eracleous2001NGCLines}, as do three of the changing-look LINERS found via follow-up of variability-selected changing-look AGN candidates from optical time-domain survey data \citep{Frederick2019ALINERs}. A double-peaked profile appeared in nuclear transient ZTF19aagwzod after an X-ray flare and the onset of strong optical variability \citep{Frederick2020AX-rays}. AT2017bcc was discovered to have time-varying double-peaked emission from an evolving accretion disk following an optical flare in a previously quiescent and inactive galaxy \citep{Ridley2023Time-varyingAT2017bcc}. J0950+5128, an AGN observed to switch from radio-quiet to radio-loud in VLASS \citep{Nyland2020QuasarsFIRST,Zhang2022TransientSurveys}, also exhibits evidence for an evolving disk profile \citep{Breiding2021TheArray}. 

Some members of other unusual AGN sub-populations, discovered in time-domain surveys, also exhibit double-peaked Balmer profiles. Five of the AGN which were found to be spatially offset from their host galaxy nuclei in ZTF imaging were DPEs \citep{Ward2021AGNsFacility}. A double-peaked profile was also observed in SDSSJ1430+2303 (ZTF18aarippg). \citet{Jiang2022Tick-Tock:Binary} attributed periodic-like features in the optical light curve of SDSSJ1430+2303 to the decaying orbit of an SMBH binary with a highly eccentric orbit and uneven mass ratio. The double-peaked profile showed a change in the relative strength of the blue and red peaks compared to 10 year old archival data and this was interpreted as the orbital motion of two broad-line regions in the binary. However, continued optical monitoring supported an alternative hypothesis of a single, rotating accretion disk \citep{Dotti2022OpticalCandidate}, suggesting that the profile evolution was due to the more typical disk evolution processes observed in DPEs. NICER monitoring of SDSSJ1430+2303 revealed the presence of quasi-periodic hard X-ray flares caused by magnetic reconnection events in the corona, which may provide clues into the accretion state of the object \citep{Masterson2023UnusualJ1430+2303}. More generally, the binary supermassive black hole interpretation has been shown to be incompatible with line profiles that have well-defined double peaks \citep[][and references therein]{Doan2020AnLines}. It has also been shown that observable signatures from individual broad emission lines in SMBH binaries arise in fewer than 1 in $10^4$ AGN due to the trade-off between having separations large enough that the broad-line regions remain attached to individual AGN and small enough that orbital velocities are detectable \citep{Kelley2020ConsiderationsAGN}. Other families of models invoking biconical outflows or a spherical BLR illuminated anisotropically are also disfavored by reverberation mapping and basic physical arguments. A comprehensive summary of the arguments can be found in \citet{Eracleous2009Double-peakedNuclei} and \citet{Eracleous2003CompletionNuclei}. 

The various recent discoveries of disk emission from variability-selected AGN and other transient phenomena in time-domain surveys motivates a large-scale population analysis of optically variable double-peaked emitters. In particular, if we are able to understand the population-wide properties of the longer-lived accretion disks in AGN, we will be better placed to understand the different properties of transient disks appearing in single TDE-driven accretion episodes.  

Optical variability is also an important clue for understanding physical differences between DPEs and other broad-line AGN which are implied by other multi-wavelength signatures. Double-peaked H$\alpha$ and H$\beta$ profiles are most commonly visible in low luminosity, low-accretion rate AGN \citep{Eracleous1994Doubled-peakedNuclei,Ho2000DoublepeakedLINERs, Ho2008NuclearGalaxies}. They are also associated with radio-loud elliptical hosts with large bulge and black hole masses \citep{Eracleous1994Doubled-peakedNuclei,Eracleous2003CompletionNuclei}. DPEs are 1.6$\times$ more likely to have radio emission and 1.5$\times$ more likely to have soft X-ray emission than other broad-line AGN \citep{Strateva2003Double-peakedNuclei}. \citet{Zhang2017PropertiesLines} found that that the optical variability properties of DPEs differed from other broad-line AGN, having damped random walk (DRW) characteristic timescales $2.7\times$ larger than a control sample in CSS and SDSS light curves. 

Providing an explanation for the differences between DPEs and other broad-line AGN while accounting for these many population differences is challenging. \citet{Storchi-Bergmann2017Double-PeakedNuclei} predict that double peaked profiles are ubiquitous in broad-line AGN, but are only observed when the inclination angle is $\gtrsim20$\textdegree \ so that the separate peaks of the accretion disk are observable, but $\lesssim37$\textdegree \ so that the accretion disk emission is not blocked by the obscuring torus. This would result in an observed $\sim60$\% fraction of broad-line AGN with double peaks, but other factors may reduce this fraction. For example, if the AGN is in a high accretion state, contributions from gas that is not part of the disk (e.g., gas that is outflowing or accelerating away from the disk) is expected to dominate the  broad-line emission and fill the dip between the peaks \citep{Storchi-Bergmann2017Double-PeakedNuclei}. 

Some broad-line AGN emission models invoke two phases: outflowing sub-critical density gas produces broad Gaussian lines, while supercritical gas close to the disk surface produces double-peaked emission.  These models may explain the transition between Type I and true Type II AGN and may also explain the higher incidence of double-peaked broad Balmer profiles in low-luminosity AGN \citep{Popovic2004ContributionModel,Bon2009ContributionNuclei, LaMura2009BalmerNuclei, Elitzur2014EvolutionNuclei}. Disk-wind AGN models may also explain the higher luminosity of double-peaked structures relative to broad line gas in low-luminosity AGN compared to standard Seyfert 1 nuclei \citep{Elitzur2014EvolutionNuclei,Storchi-Bergmann2017Double-PeakedNuclei}. These models are supported by reverberation mapping of Seyfert 1 nuclei showing that even when the H$\beta$ profiles are not double peaked, the root mean square taken across time series of H$\beta$ spectra still often shows a double-peaked profile, implying that the most variable, innermost broad line gas is in a disk \citep{Denney2010ReverberationGalaxies, Schimoia2017EvolutionNGC7213,Storchi-Bergmann2017Double-PeakedNuclei}.

In this paper we present, for the first time, a large-scale population study of disk-emitting AGN with both comprehensive time-domain and spectroscopic analysis. For our variability analysis we have used observations from the Zwicky Transient Facility \citep[ZTF;][]{Bellm2019,Graham2019,Dekany2020TheSystem}, an ongoing optical survey which began in March 2018 and achieves single epoch limiting magnitudes of $\sim21$ in the g-, r- and i-bands over a survey footprint of 23,675 deg$^2$. For our sample of 250 DPEs, we present both power spectrum analysis of ZTF light curves and spectroscopically derived disk geometries from fits to the double-peaked line profiles.

\begin{figure*}
\gridline{\fig{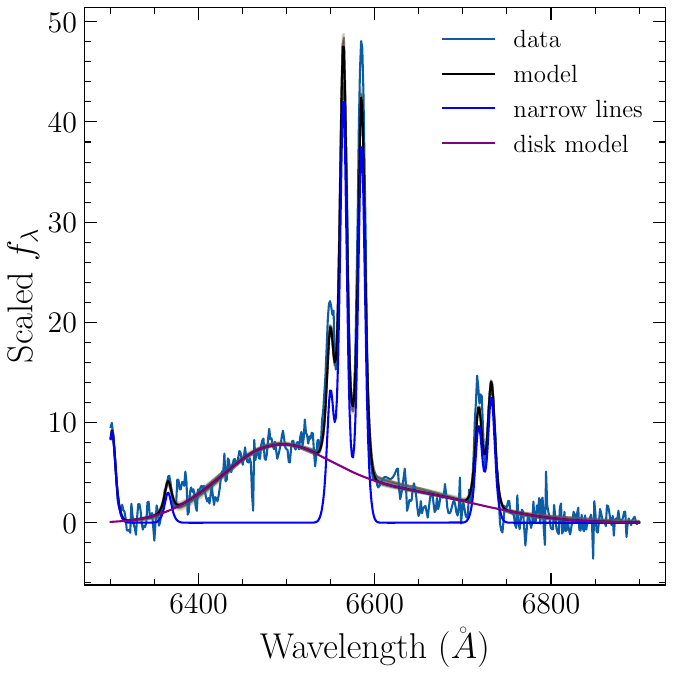}{0.24\textwidth}{a) ZTF18aarippg} 
 \fig{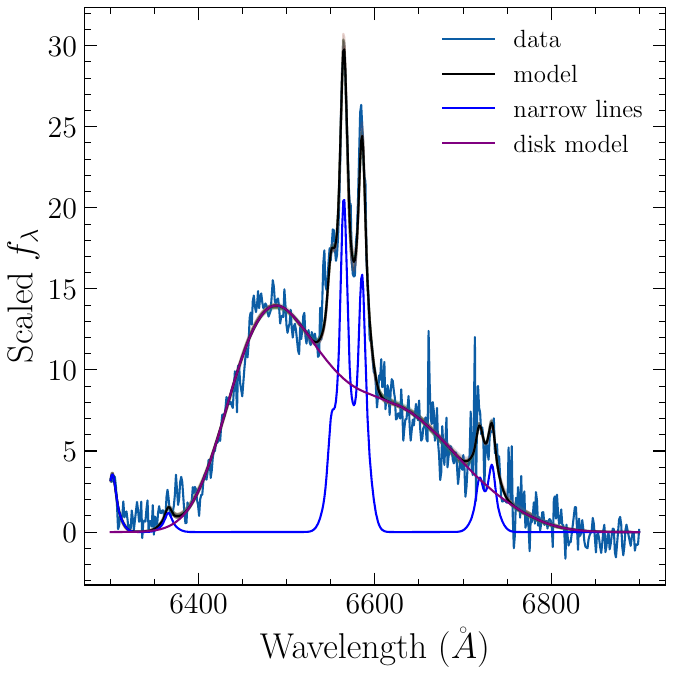}{0.24\textwidth}{b) ZTF19aatllxr}\fig{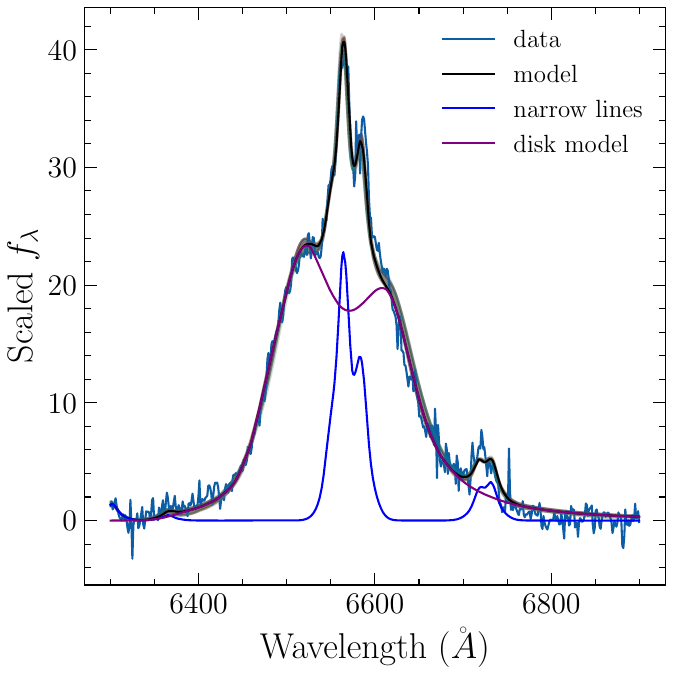}{0.24\textwidth}{c) ZTF18acyygbx} \fig{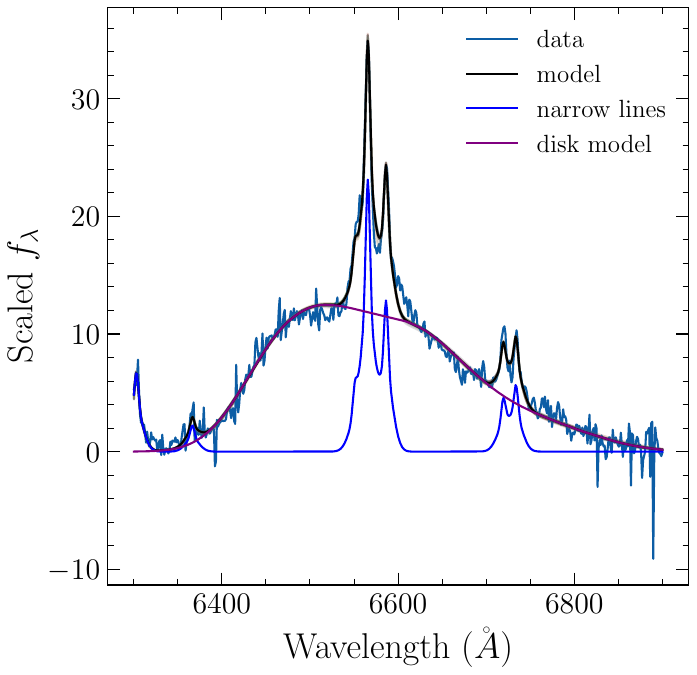}{0.24\textwidth}{d) ZTF19aadgigp} }
 \gridline{\fig{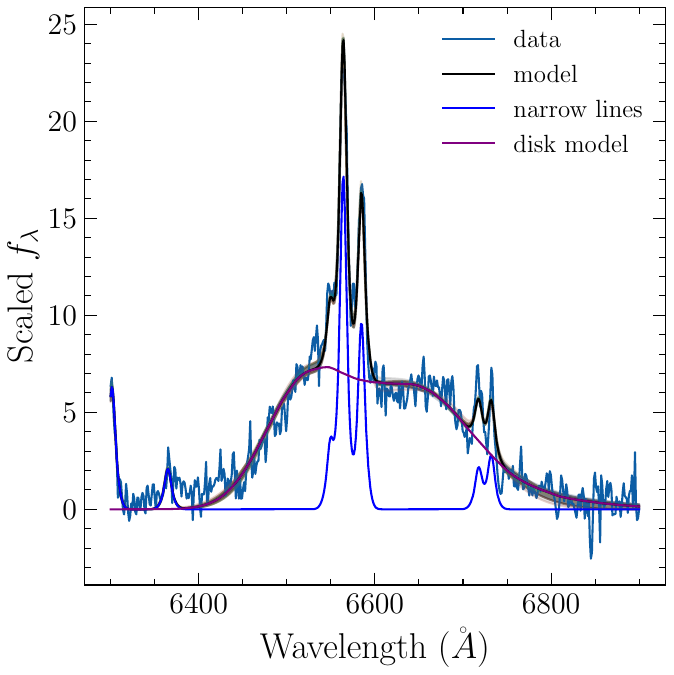}{0.24\textwidth}{e) ZTF18achchge}
\fig{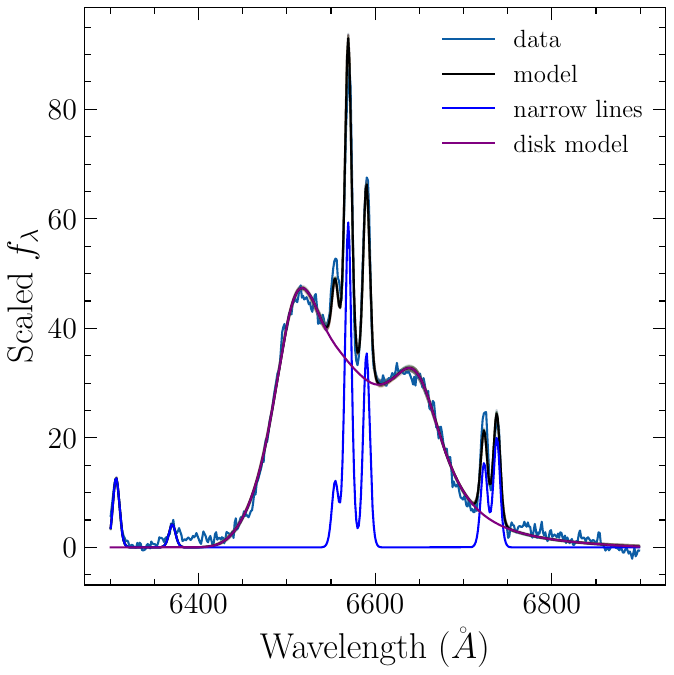}{0.24\textwidth}{f) ZTF18aahiqst}\fig{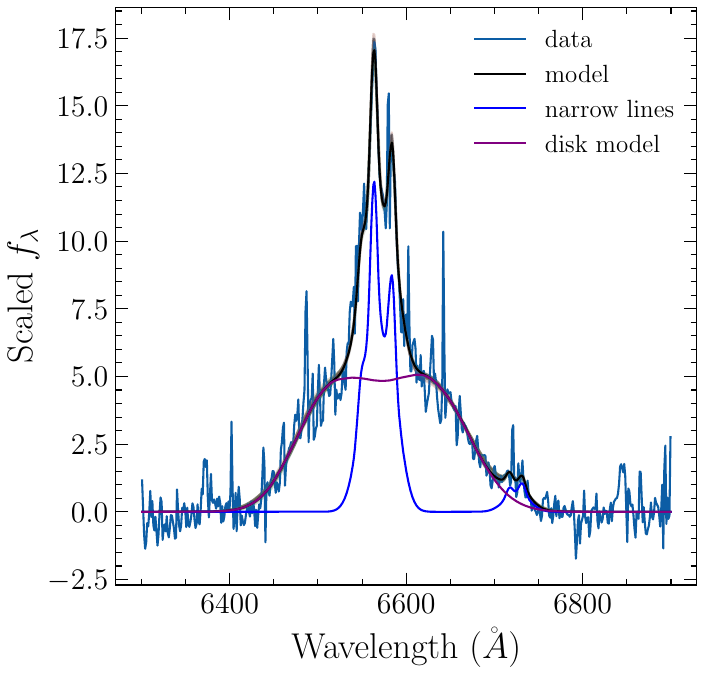}{0.24\textwidth}{g) ZTF19acdtzdd}\fig{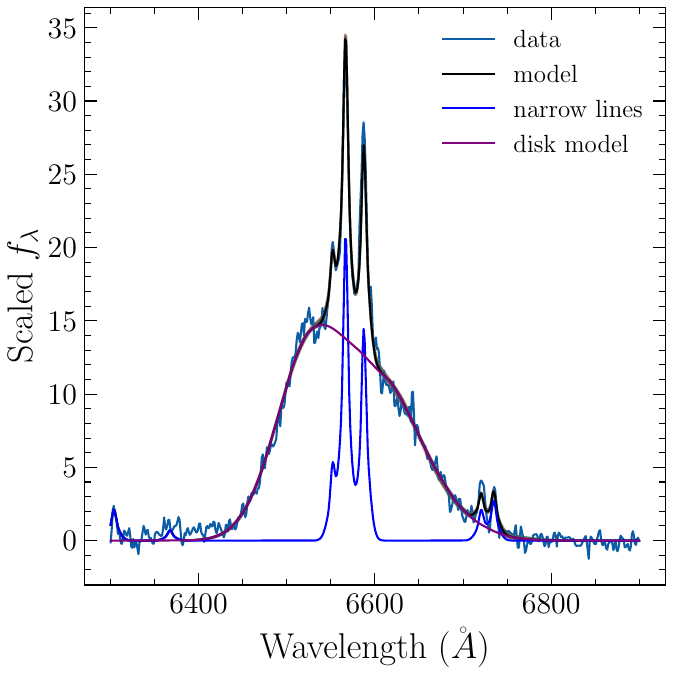}{0.24\textwidth}{h) ZTF18accdmzk}}
\gridline{ \fig{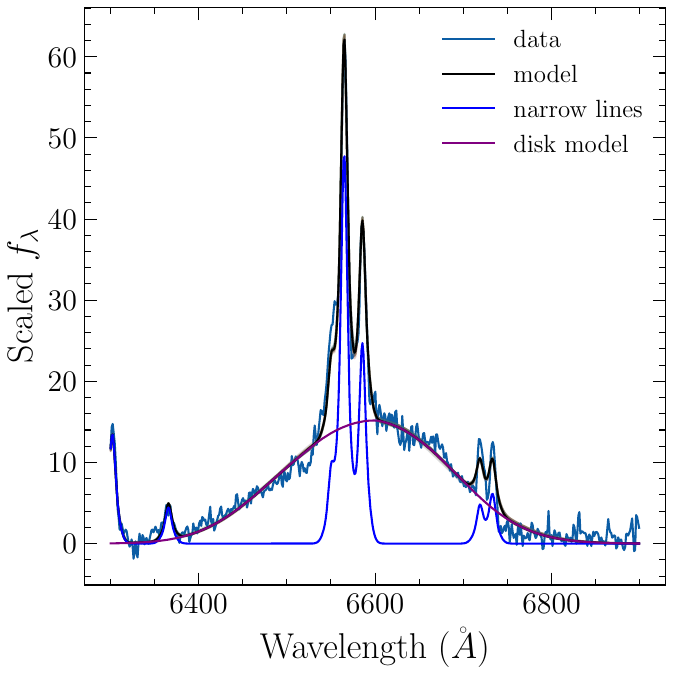}{0.24\textwidth}{i) ZTF19aawlmtf} \fig{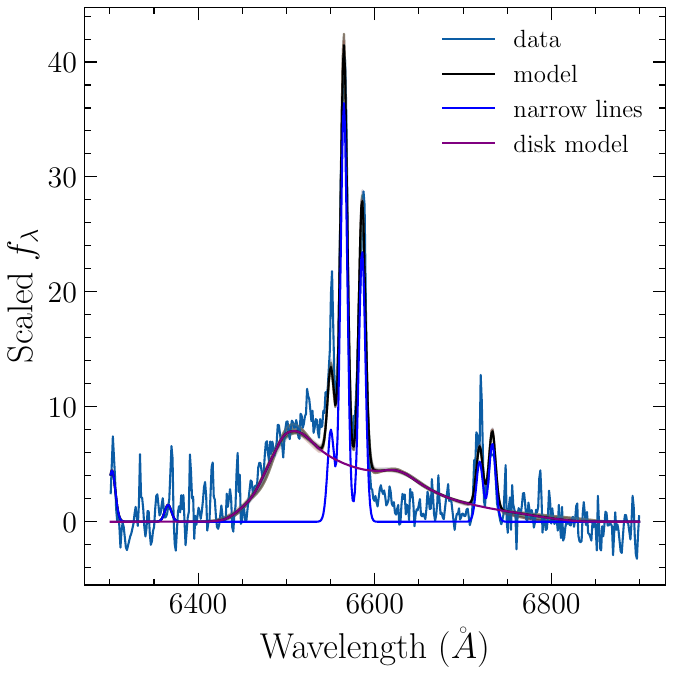}{0.24\textwidth}{j) ZTF18accwjao}\fig{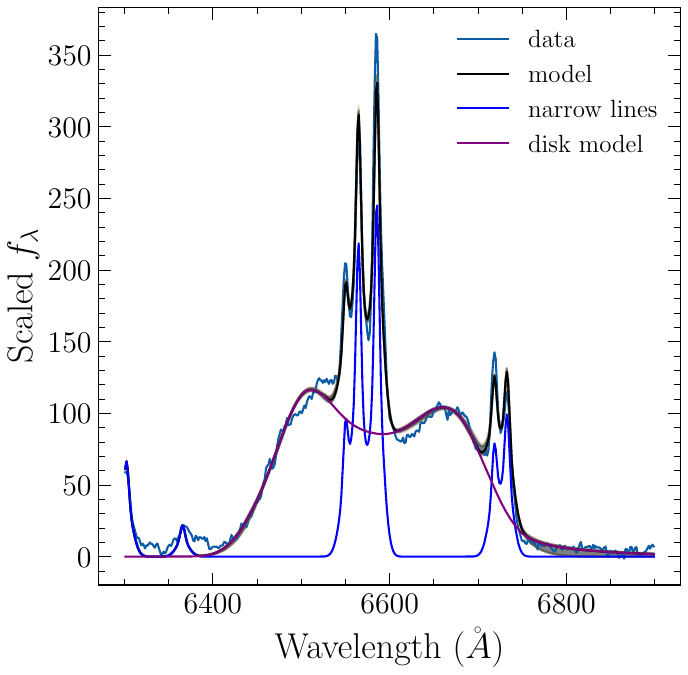}{0.24\textwidth}{k) ZTF18aarywbt}\fig{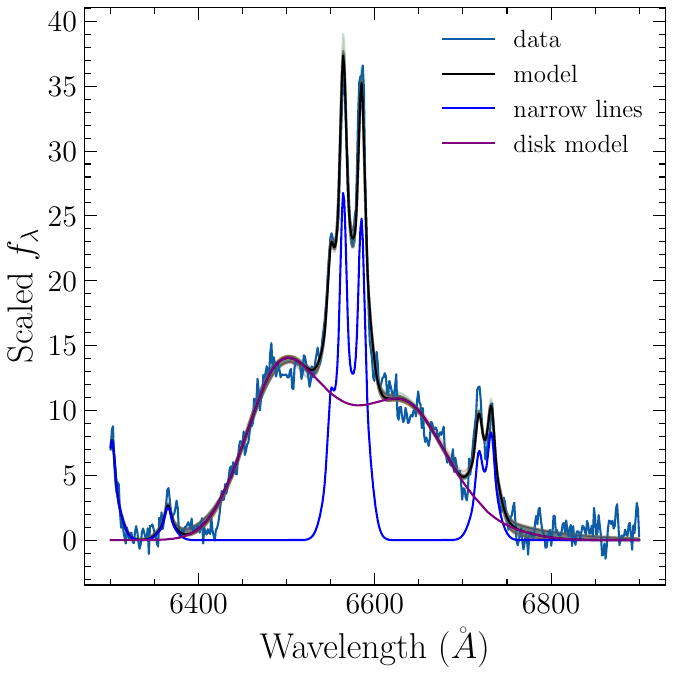}{0.24\textwidth}{l) ZTF18aaqgbag}}
\gridline{\fig{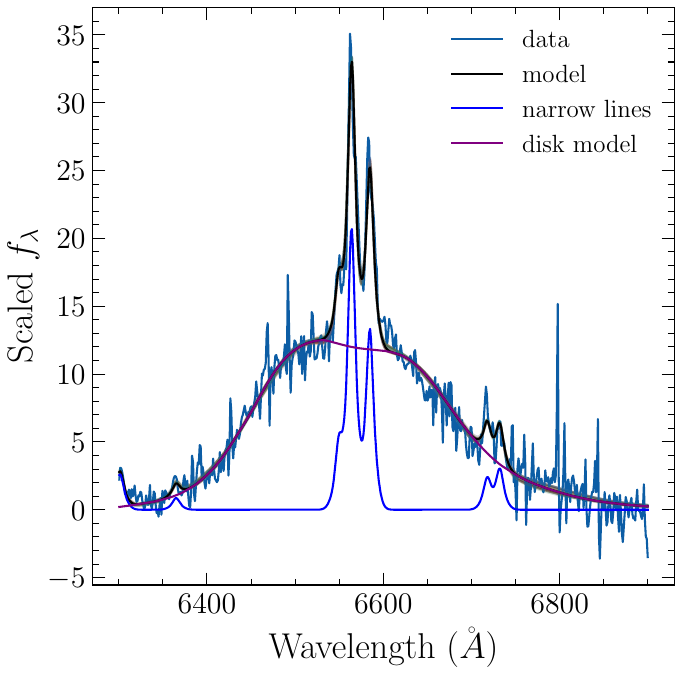}{0.24\textwidth}{m) ZTF18aaqdmih} \fig{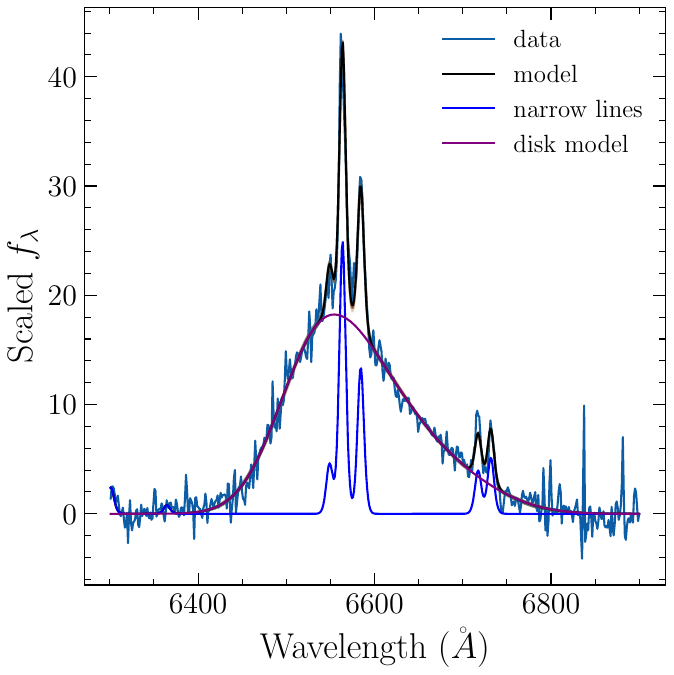}{0.24\textwidth}{n) ZTF19abizomu}\fig{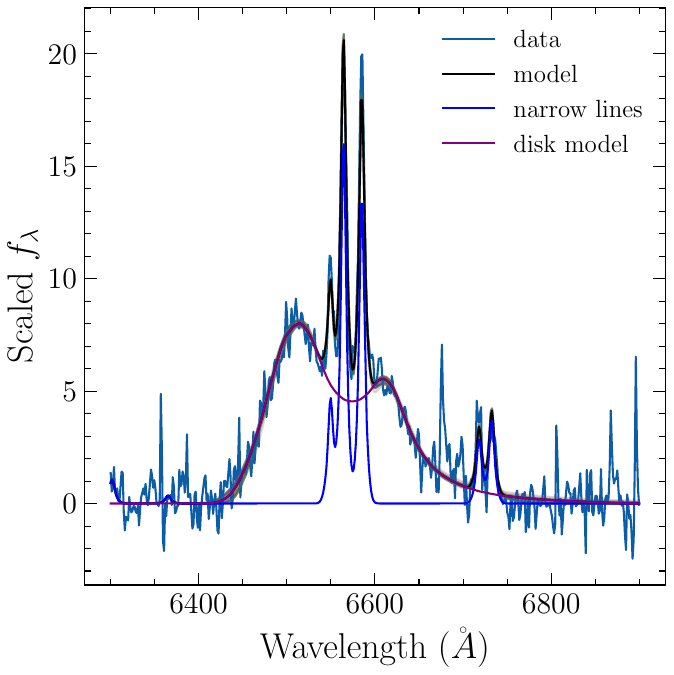}{0.24\textwidth}{o) ZTF19aawfvxi}\fig{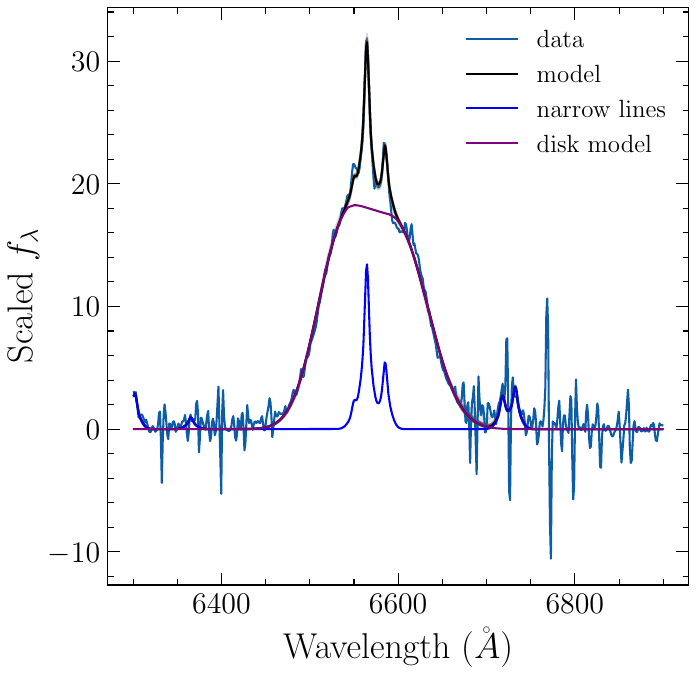}{0.24\textwidth}{p) ZTF18aaznjgn}}
\caption{Examples of different disk profile morphologies for 16 DPEs identified in the parent sample. We include the profiles of 3 DPEs with radio jet imaging in Figure \ref{fig:radioimages} (ZTF18aaqdmih (m), ZTF18aarywbt (k) and ZTF19abizomu(n)) and DPEs with notable light curves presented in Figure \ref{fig:lcs_notable} (ZTF18aarippg (a) and ZTF18aaznjgn (p)). We note that in rare cases like ZTF18accwjao (j), the relative strengths of the blue and red peaks could not be fully accounted for when the spiral arm contrast ratio was restricted to $<8$. When the parameters describing internal structure within the disk were more flexible, a better model was found.}
\label{fig:profile_egs}
\end{figure*}

There are three primary goals to this paper. Firstly, we aim to determine how common DPEs are amongst variable AGN -- and what fraction of DPEs show time-evolution in their broad-line profiles -- in order to see if the time-evolving broad-line profiles of TDEs and SMBH binary candidates reported in the literature are unusual compared to optically variable AGN. Secondly, we aim to determine if the optical/IR variability properties, and the radio properties, of the DPE population are any different to other broad-line AGN, in order to see if the differences between the two populations are more consistent with viewing angle effects or with a difference in accretion state. Finally, we aim to provide a catalog of optical light curve power spectrum properties, spectroscopically-derived accretion disk geometries (including key parameters such as inclination angle), and radio emission properties to inform future analysis of AGN and TDE accretion disks and jets, and the dependence of AGN properties on inclination angle. 

In Section 2 we describe the selection of optically variable broad-line AGN in ZTF and the spectroscopic criteria used to identify the sub-sample of DPEs. We present fitting of the spectra of the double-peaked profiles with accretion disk models and provide a catalog of accretion disk properties for the 250 DPEs. In Section 3 we characterize the optical variability properties of the DPEs compared to the control sample of broad-line AGN and present examples of notable light curves from the DPE population. In Section 4 we analyze the radio properties of the DPEs and present 20-34 cm imaging of radio jets for 3 objects. In Section 5 we present spectroscopic monitoring a DPE sub-sample, showing that 50\% exhibit large changes in the relative strengths of the red and blue peaks over decade-long timescales. We also discuss the unusual properties of two objects which are atypical for disk emitters and may be worthy of additional follow-up. In Section 6 we discuss the overall accretion disk and variability properties of the DPE and AGN populations in light of other transients with visible accretion disks, as well as the search for SMBH binary candidates in time-domain data. We summarize our conclusions in Section 7.

\section{Spectroscopic classification of DPEs}
\subsection{Selection of variable broad-line AGN with SDSS spectroscopy}
To produce a parent sample of 1549 optically variable broad-line AGN we started with the 5000 variable AGN identified in ZTF time-domain survey data in \citet{Ward2021AGNsFacility} and required the AGN to have redshifts $z<0.4$ and to have an increase in flux of 1.5 magnitudes in either the g- or r-band difference image photometry and to be classified as GALAXY AGN BROADLINE' or `QSO BROADLINE' in the SDSS DR14 spectroscopic catalog \citep{Blanton2017SloanUniverse}.

\begin{deluxetable*}{llllccccccc}
\tabletypesize{\scriptsize}
\tablecolumns{10}
\tablewidth{0pt}
\tablecaption{Properties of the 250 DPE candidates from ZTF (full table available online) \label{table:ztfcands}}
\tablehead{
\colhead{ZTF ID} & \colhead{RA} & \colhead{Dec} & \colhead{z} &\colhead{Log Amp.}& \colhead{PL index} &\colhead{Turn.}&\colhead{VLASS E1}&\colhead{VLASS E2}&\colhead{RACS} \\[-0.3cm]
\colhead{} & \colhead{(hms)} & \colhead{(dms)} & \colhead{} &\colhead{}& \colhead{} &\colhead{(yr$^{-1}$)}&\colhead{(mJy)}&\colhead{(mJy)}&\colhead{(mJy)}}
\startdata
ZTF18aaaotwe&13:13:10.376&15:45:03.347&$0.0657$&$-2.8^{+0.3}_{-0.3}$&$4.2^{+1.7}_{-2.4}$&$3.2^{+1.6}_{-2.0}$&ND&ND&ND\\
ZTF18aaaovpz&10:39:13.820&09:40:02.779&$0.217$&$-2.8^{+0.4}_{-0.7}$&$2.2^{+3.3}_{-1.0}$&$0.3^{+1.1}_{-0.3}$&ND&ND&ND\\
ZTF18aaavwka&12:35:44.246&16:05:35.978&$0.0711$&$-2.5^{+0.0}_{-0.1}$&$2.5^{+3.4}_{-2.4}$&$68.0^{+28.1}_{-46.4}$&ND&ND&ND\\
ZTF18aabkubl&11:25:58.744&20:05:54.825&$0.133$&$-1.2^{+0.0}_{-0.1}$&$3.9^{+1.9}_{-2.5}$&$4.2^{+0.7}_{-2.1}$&$469.95\pm3.08$&$469.95\pm3.08$&ND\\
ZTF18aabylvn&14:17:59.554&25:08:12.590&$0.0163$&$-1.4^{+0.0}_{-0.0}$&$0.0^{+0.1}_{-0.0}$&$3.3^{+1.7}_{-2.9}$&$3.21\pm0.29$&$3.21\pm0.29$&ND\\
ZTF18aacajqc&10:29:46.791&40:19:13.636&$0.0673$&$-3.2^{+0.3}_{-0.3}$&$4.0^{+1.9}_{-2.7}$&$1.1^{+1.4}_{-0.9}$&ND&ND&-\\
ZTF18aacbjdm&12:32:03.637&20:09:29.529&$0.0636$&$-1.6^{+0.1}_{-0.2}$&$2.4^{+2.5}_{-2.0}$&$4.2^{+0.8}_{-2.6}$&ND&ND&$3.6\pm1.07$\\
ZTF18aaccaxc&13:37:39.948&39:09:16.941&$0.0198$&$-3.3^{+0.1}_{-0.2}$&$1.1^{+3.3}_{-0.6}$&$0.8^{+3.9}_{-0.8}$&$1.54\pm0.27$&$1.54\pm0.27$&-\\
ZTF18aacckko&14:05:44.376&40:51:16.676&$0.0664$&$-2.0^{+0.3}_{-0.2}$&$3.0^{+2.8}_{-1.6}$&$1.8^{+2.2}_{-1.6}$&ND&ND&-\\
ZTF18aacddjc&13:42:20.173&38:42:09.520&$0.0788$&$-3.7^{+0.4}_{-0.5}$&$3.3^{+2.7}_{-2.6}$&$3.2^{+92.7}_{-2.8}$&ND&ND&-\\
ZTF18aacdpbi&09:05:14.486&41:51:53.493&$0.1764$&$-2.7^{+0.3}_{-0.2}$&$1.5^{+0.8}_{-0.4}$&$0.1^{+0.7}_{-0.1}$&ND&ND&-\\
ZTF18aacdvjp&09:37:28.578&32:45:48.310&$0.127$&$-2.2^{+0.3}_{-0.3}$&$2.1^{+3.2}_{-0.9}$&$0.9^{+2.5}_{-0.8}$&ND&ND&-\\
ZTF18aachojf&08:39:49.670&48:47:01.667&$0.0392$&$-2.4^{+0.0}_{-0.1}$&$4.7^{+1.3}_{-2.6}$&$3.9^{+1.0}_{-2.0}$&ND&ND&-\\
ZTF18aacrkse&09:05:14.481&41:51:53.825&$0.176$&$-2.7^{+0.3}_{-0.3}$&$1.7^{+2.8}_{-0.7}$&$0.5^{+2.5}_{-0.4}$&ND&ND&-\\
ZTF18aadgbva&09:11:13.384&40:01:11.238&$0.201$&$-2.5^{+0.3}_{-0.2}$&$3.2^{+2.6}_{-1.6}$&$1.7^{+2.2}_{-1.5}$&ND&ND&-\\
ZTF18aaercku&09:19:10.523&25:49:53.960&$0.366$&$-2.9^{+0.3}_{-0.3}$&$2.8^{+3.1}_{-1.4}$&$0.8^{+1.7}_{-0.8}$&ND&ND&ND\\
ZTF18aahfere&10:38:53.307&39:21:51.218&$0.0548$&$-2.5^{+0.3}_{-0.3}$&$2.4^{+3.3}_{-0.9}$&$0.4^{+1.0}_{-0.4}$&$1.08\pm0.24$&$1.08\pm0.24$&-\\
ZTF18aahfhsm&10:19:06.786&23:18:37.839&$0.0646$&$-2.2^{+0.3}_{-0.3}$&$2.5^{+3.4}_{-1.6}$&$1.8^{+2.8}_{-1.8}$&ND&ND&ND\\
ZTF18aahfohe&12:16:07.085&50:49:30.174&$0.031$&$-2.1^{+0.3}_{-0.2}$&$1.7^{+1.6}_{-0.6}$&$0.4^{+1.8}_{-0.4}$&$3.46\pm0.25$&$3.46\pm0.25$&-\\
ZTF18aahfssj&12:30:59.742&35:45:42.828&$0.1004$&$-1.4^{+0.1}_{-0.2}$&$4.4^{+1.5}_{-2.9}$&$3.1^{+1.4}_{-2.5}$&ND&ND&-\\
ZTF18aahgojc&13:37:39.821&39:09:16.036&$0.0198$&$-3.6^{+0.3}_{-0.6}$&$1.4^{+3.6}_{-0.9}$&$1.2^{+3.6}_{-1.1}$&$1.54\pm0.27$&$1.54\pm0.27$&-\\
ZTF18aahhuol&11:37:24.523&35:09:12.619&$0.263$&$-3.1^{+0.3}_{-0.3}$&$3.8^{+2.1}_{-1.7}$&$0.8^{+0.9}_{-0.7}$&ND&ND&-\\
ZTF18aahhvqh&12:26:30.999&25:25:22.045&$0.134$&$-2.5^{+0.3}_{-0.3}$&$2.4^{+2.7}_{-0.9}$&$0.7^{+1.8}_{-0.7}$&ND&ND&ND\\
ZTF18aahiqst&11:03:40.320&37:29:25.080&$0.0745$&$-3.0^{+0.3}_{-0.3}$&$4.3^{+1.6}_{-2.7}$&$0.7^{+0.7}_{-0.7}$&$2.15\pm0.24$&$2.15\pm0.24$&-\\
\enddata
\vspace{0.1cm}
\tablecomments{Properties of the 250 variable DPE candidates in ZTF. Objects in the table are arranged in lexigraphical order by ZTF name. Spectroscopic redshifts from SDSS are shown in the 4th column. In columns 5-7 we show the amplitude, power law index and turnover frequency from modeling of the power spectrum derived from the g-band ZTF light curve (see Section 3). In columns 8-10 we show the the radio fluxes for epoch 1 (2017-2018) and epoch 2 (2019-2021) of the 20cm VLASS survey and for epoch 1 of the 34cm RACS-low survey. Non-detections are indicated by ND, and a dash indicates that the source was not within the surveyed region.}
\end{deluxetable*}

\begin{deluxetable*}{llllllllll}
\tabletypesize{\scriptsize}
\tablecolumns{11}
\tablewidth{0pt}
\tablecaption{Best-fit accretion disk parameters (full table available online). \label{table:diskparams}}
\tablehead{\colhead{ZTF ID} &\colhead{$\xi_{1}$}&\colhead{$\xi_{2}$}&\colhead{$\sigma$ (km/s)}&\colhead{$i$ (deg)}&\colhead{$q$}&\colhead{$w$ (deg)}&\colhead{$A_s$}&\colhead{$\psi$ (deg)}&\colhead{$\phi$ (deg)}} 
\startdata
ZTF18aaaotwe&$160^{+20}_{-40}$&$2350^{+1140}_{-1100}$&$2220^{+140}_{-110}$&$12^{+1}_{-1}$&$2.0^{+0.3}_{-0.3}$&$70^{+10}_{-7}$&$7^{+1}_{-1}$&$40^{+10}_{-10}$&$330^{+40}_{-40}$\\
ZTF18aaaovpz&$210^{+70}_{-140}$&$1770^{+200}_{-250}$&$1040^{+300}_{-430}$&$30^{+2}_{-1}$&$1.2^{+0.3}_{-1.0}$&$59^{+6}_{-6}$&$7^{+1}_{-1}$&$310^{+10}_{-10}$&$190^{+10}_{-10}$\\
ZTF18aaavwka&$310^{+100}_{-100}$&$2550^{+40}_{-40}$&$230^{+60}_{-70}$&$18^{+1}_{-1}$&$1.1^{+0.2}_{-0.3}$&$62^{+14}_{-13}$&$1^{+1}_{-1}$&$50^{+30}_{-10}$&$320^{+60}_{-320}$\\
ZTF18aabkubl&$180^{+50}_{-50}$&$310^{+70}_{-80}$&$670^{+120}_{-120}$&$13^{+1}_{-1}$&$1.6^{+0.5}_{-0.6}$&$55^{+19}_{-16}$&$5^{+1}_{-1}$&$30^{+20}_{-20}$&$290^{+30}_{-20}$\\
ZTF18aabylvn&$210^{+10}_{-20}$&$3980^{+20}_{-10}$&$1400^{+20}_{-10}$&$19^{+1}_{-1}$&$1.7^{+0.0}_{-0.0}$&$70^{+2}_{-2}$&$7^{+1}_{-1}$&$330^{+10}_{-10}$&$140^{+10}_{-10}$\\
ZTF18aacajqc&$130^{+20}_{-10}$&$3380^{+420}_{-390}$&$590^{+60}_{-50}$&$19^{+1}_{-1}$&$0.9^{+0.1}_{-0.1}$&$69^{+7}_{-6}$&$7^{+1}_{-1}$&$300^{+10}_{-10}$&$200^{+10}_{-10}$\\
ZTF18aacbjdm&$220^{+10}_{-10}$&$760^{+70}_{-60}$&$1450^{+10}_{-10}$&$8^{+1}_{-1}$&$2.4^{+0.1}_{-0.1}$&$77^{+3}_{-2}$&$7^{+1}_{-1}$&$310^{+10}_{-10}$&$250^{+10}_{-10}$\\
ZTF18aaccaxc&$160^{+30}_{-50}$&$3820^{+10}_{-10}$&$940^{+120}_{-120}$&$11^{+1}_{-1}$&$1.9^{+0.2}_{-0.2}$&$69^{+19}_{-8}$&$6^{+1}_{-1}$&$40^{+40}_{-20}$&$350^{+10}_{-10}$\\
ZTF18aacckko&$200^{+20}_{-30}$&$3420^{+10}_{-10}$&$1620^{+0}_{-0}$&$21^{+1}_{-1}$&$1.9^{+0.1}_{-0.1}$&$41^{+22}_{-28}$&$3^{+2}_{-2}$&$360^{+10}_{-10}$&$240^{+10}_{-230}$\\
ZTF18aacddjc&$320^{+80}_{-90}$&$2090^{+20}_{-30}$&$1620^{+260}_{-190}$&$17^{+1}_{-1}$&$1.4^{+0.4}_{-0.6}$&$33^{+19}_{-31}$&$3^{+2}_{-2}$&$360^{+40}_{-50}$&$30^{+70}_{-40}$\\
ZTF18aacdpbi&$190^{+30}_{-40}$&$2430^{+50}_{-50}$&$1870^{+60}_{-50}$&$13^{+1}_{-1}$&$1.3^{+0.2}_{-0.2}$&$75^{+6}_{-4}$&$7^{+1}_{-1}$&$320^{+10}_{-10}$&$210^{+30}_{-30}$\\
ZTF18aacdvjp&$170^{+10}_{-10}$&$3900^{+150}_{-70}$&$1250^{+40}_{-40}$&$17^{+1}_{-1}$&$1.6^{+0.0}_{-0.0}$&$37^{+8}_{-7}$&$7^{+1}_{-1}$&$340^{+10}_{-10}$&$70^{+20}_{-20}$\\
ZTF18aachojf&$190^{+20}_{-20}$&$1010^{+20}_{-120}$&$730^{+30}_{-30}$&$11^{+1}_{-1}$&$1.4^{+0.3}_{-0.3}$&$79^{+1}_{-0}$&$8^{+1}_{-1}$&$60^{+10}_{-10}$&$290^{+10}_{-10}$\\
ZTF18aacrkse&$200^{+40}_{-50}$&$2550^{+70}_{-70}$&$1870^{+60}_{-60}$&$13^{+1}_{-1}$&$1.4^{+0.2}_{-0.3}$&$74^{+6}_{-4}$&$7^{+1}_{-1}$&$320^{+10}_{-10}$&$210^{+20}_{-20}$\\
ZTF18aadgbva&$250^{+30}_{-40}$&$3730^{+360}_{-200}$&$280^{+80}_{-80}$&$18^{+1}_{-1}$&$1.5^{+0.1}_{-0.1}$&$31^{+23}_{-34}$&$0^{+1}_{-1}$&$0^{+40}_{-40}$&$140^{+250}_{-260}$\\
ZTF18aaercku&$290^{+60}_{-90}$&$2260^{+830}_{-1220}$&$1250^{+80}_{-120}$&$11^{+1}_{-1}$&$1.9^{+0.5}_{-0.4}$&$50^{+25}_{-19}$&$6^{+3}_{-1}$&$30^{+30}_{-10}$&$220^{+10}_{-10}$\\
ZTF18aahfere&$230^{+30}_{-30}$&$1220^{+190}_{-310}$&$1250^{+70}_{-60}$&$12^{+1}_{-1}$&$1.7^{+0.5}_{-0.6}$&$38^{+6}_{-15}$&$6^{+1}_{-1}$&$320^{+10}_{-20}$&$200^{+20}_{-20}$\\
ZTF18aahfhsm&$200^{+20}_{-20}$&$1370^{+80}_{-80}$&$1000^{+50}_{-50}$&$11^{+1}_{-1}$&$1.5^{+0.3}_{-0.3}$&$78^{+4}_{-2}$&$7^{+1}_{-1}$&$50^{+10}_{-10}$&$290^{+10}_{-10}$\\
ZTF18aahfohe&$240^{+10}_{-10}$&$3960^{+70}_{-30}$&$900^{+90}_{-110}$&$23^{+1}_{-1}$&$1.7^{+0.1}_{-0.1}$&$44^{+5}_{-11}$&$7^{+1}_{-1}$&$340^{+10}_{-10}$&$70^{+10}_{-10}$\\
ZTF18aahfssj&$1880^{+170}_{-90}$&$3800^{+320}_{-140}$&$2040^{+120}_{-160}$&$15^{+5}_{-3}$&$1.5^{+0.5}_{-0.6}$&$44^{+29}_{-25}$&$2^{+1}_{-2}$&$20^{+40}_{-30}$&$290^{+20}_{-10}$\\
ZTF18aahgojc&$190^{+40}_{-30}$&$3820^{+10}_{-10}$&$510^{+290}_{-460}$&$14^{+3}_{-2}$&$1.9^{+0.2}_{-0.2}$&$35^{+15}_{-34}$&$7^{+1}_{-1}$&$310^{+10}_{-100}$&$30^{+140}_{-210}$\\
ZTF18aahhuol&$130^{+10}_{-10}$&$1000^{+10}_{-10}$&$1470^{+1020}_{-1050}$&$53^{+4}_{-4}$&$1.8^{+0.7}_{-0.6}$&$37^{+27}_{-35}$&$6^{+1}_{-1}$&$350^{+40}_{-60}$&$320^{+80}_{-250}$\\
ZTF18aahhvqh&$110^{+30}_{-80}$&$1930^{+10}_{-10}$&$880^{+0}_{-10}$&$17^{+1}_{-1}$&$0.8^{+0.0}_{-0.5}$&$14^{+6}_{-10}$&$0^{+1}_{-1}$&$360^{+10}_{-10}$&$360^{+10}_{-80}$\\
ZTF18aahiqst&$100^{+20}_{-10}$&$1310^{+40}_{-50}$&$750^{+40}_{-40}$&$21^{+1}_{-1}$&$0.8^{+0.0}_{-0.0}$&$68^{+6}_{-6}$&$3^{+1}_{-1}$&$300^{+10}_{-10}$&$190^{+10}_{-10}$\\
\enddata
\vspace{0.1cm}
\tablecomments{Best-fit disk parameters from modeling the H$\alpha$ broad-line regions of the AGN with the circular accretion disk model from \cite{Chen1989}. We show the following H$\alpha$ disk parameters: inner radius $\xi_{1}$ (gravitational radii), outer radius $\xi_{2}$ (gravitational radii), turbulent broadening $\sigma$ (km/s), inclination angle $i$ (deg), spiral arm width $w$ (deg), spiral arm amplitude expressed as contrast ratio $A_s$, spiral arm pitch angle $\psi$ (deg) and spiral arm phase $\phi$ (deg).}
\end{deluxetable*}

In order to find the AGN with double-peaked broad lines amongst the sample of 1549 broad-line AGN, we modeled the archival SDSS spectra of the AGN. We first used Penalized Pixel Fitting (pPXF) \citep{Cappellari2003,Cappellari2017ImprovingFunctions} to model and subtract the stellar continuum and absorption lines. The continuum-subtracted spectra of the full AGN sample have been made available in a github repo containing the intermediate data products\footnote{https://github.com/charlotteaward/ZTF-DPEs}. 

\subsection{Fitting of broad H$\alpha$ profiles with circular accretion disk models}
After continuum subtraction we modeled the H$\alpha$ broad-line region of each AGN, regardless of whether there were discernible double peaks or shoulders, with the circular accretion disk model from \cite{Chen1989}. We fit a circular accretion disk model describing the H$\alpha$ broad emission line regions combined with the narrow emission lines from H$\alpha$, [S\,{\sc ii}] $\lambda$6717, 6731, [N\,{\sc ii}] $\lambda$6550, 6575, and [O\,{\sc i}] $\lambda$6302, 6366. The circular accretion disk model was chosen over an elliptical accretion disk model because circular models with rotating spiral arms have been shown to better reproduce the timescale of profile variability observed in DPEs compared to precessing elliptical disks \citep{Eracleous2009Double-peakedNuclei}.

The disk models had the following free parameters: the inclination angle $i$ (deg) where 0\textdegree\ is face-on and 90\textdegree\ is edge-on, a local turbulent broadening parameter $\sigma$\ (km/s), the emissivity power law index $q$, and the inner and outer dimensionless gravitational radii of the disk $\xi_1$ and $\xi_2$. We enabled a single spiral arm with free parameters amplitude $A_s$ (expressed as a contrast ratio relative to the rest of the disk), orientation angle $\phi$ (deg), width $w$ (deg), and pitch angle $\psi$ (deg). This was required to describe the flux ratio of the red and blue shoulders being $>$1 in a fraction of spectra, which has been commonly observed in other disk emitters \citep[e.g.][]{StorchiBergmann2003EvolutionHole}. We applied the following bounds on some parameters via a uniform prior: $\xi_1>50$, $w<80$, $0<\psi<60$, and $A_s<8$, based on typical parameters found for DPEs with detailed spiral arm modeling of multi-epoch spectra \citep[e.g.][]{Schimoia2012ShortNGC1097,Schimoia2017EvolutionNGC7213}. We did not include a disk wind \citep{Murray1996Wind-dominatedStars,Flohic2012EFFECTSNUCLEI, Nguyen2018EmissionLines, Chajet2013MagnetohydrodynamicDistributions} in the models because the circular disk with a single spiral arm adequately described all spectra.

The disk model was fitted simultaneously with a model for the forbidden narrow emission lines overlapping the H$\alpha$ broad-line region. The [N\,{\sc ii}], [S\,{\sc ii}], [O\,{\sc i}] doublet flux ratios were fixed to theoretical values of 2.95, 1.3, and 0.33 respectively. The narrow lines were described by two component Gaussians of the same central wavelength with 3 free parameters which were common for all narrow lines: the width of the first Gaussian component $\sigma_1$, the width of the second Gaussian component $\sigma_2$, and the flux ratio of the two components $f_1/f_2$. The amplitudes of the spectral lines are linear parameters, and so for computational expediency, we used a profile likelihood technique in which, for a given set of narrow line, broad line, and disk model parameters, we determined the amplitudes via least squares optimization. 

We first found a reasonable initial fit using the nonlinear least-squares optimisation implemented in \textsc{Python} using the \textsc{scipy} package. We then explored the posteriors using \textsc{emcee} \citep{Foreman-Mackey2013ttemcee/ttHammer} with 60 walkers initialized at the best-fit values from the least-squares fit, distributed according to the 1$\sigma$ error found from the least-squares covariance matrix. For each spectrum, the emcee fitting was run for 2400 iterations with a burn-in time of 1800 iterations. 

\begin{figure*}
\gridline{\fig{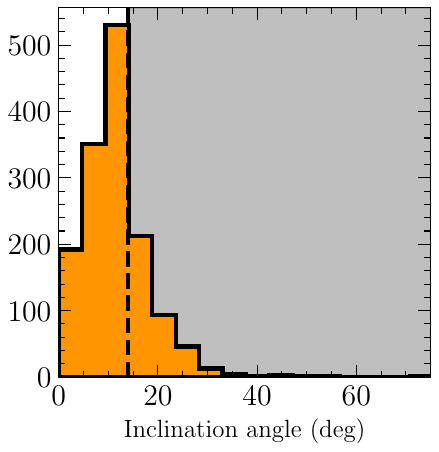}{0.32\textwidth}{} \fig{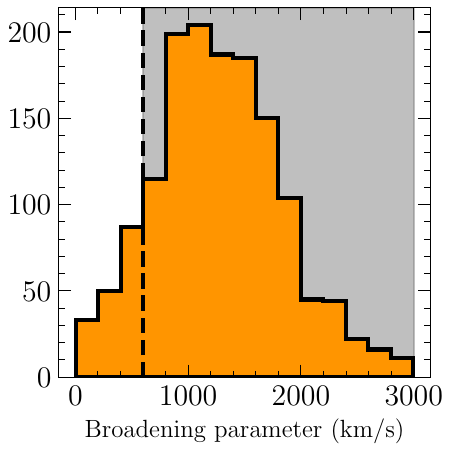}{0.33\textwidth}{} \fig{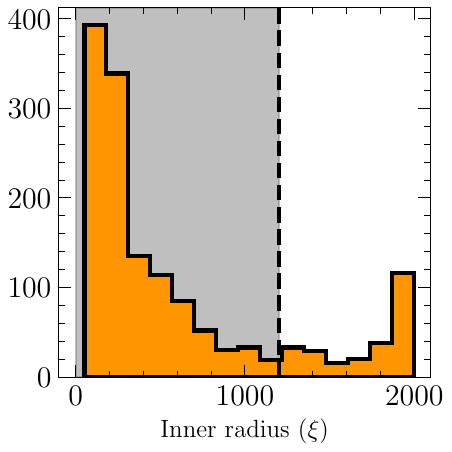}{0.33\textwidth}{}}
\caption{Histograms showing the distributions of the three disk parameters used to separate DPEs from AGN with `normal' broad lines. DPEs were classified as those with inclination angle $>14$ \textdegree, turbulent broadening $>600$ km/s, and inner radius $<1200$ gravitational radii.}
\label{fig:disksplit}
\end{figure*}

\subsection{Results from disk model fitting: profile shapes and DPE fractions}
We produced disk models for 1302 out of the 1549 AGN: the remainder had H$\alpha$ regions too strongly dominated by the narrow emission lines to produce reliable fits to the broad-line component. The H$\alpha$ broad lines for all AGN in the sample were well-described by our disk$+$narrow line model. Examples of double-peaked H$\alpha$ disk profiles and their fits are shown in Figure \ref{fig:profile_egs}. We note that the broad lines and corresponding disk models show a variety of shapes, including well-separated blue and red peaks (e.g. ZTF18aahiqst: Figure \ref{fig:profile_egs}f), a large blue-red shoulder flux ratio such that they appear to have a single velocity-offset broad-line (ZTF18accwjao: Figure \ref{fig:profile_egs}j), closely separated peaks which still require a dip in the center to correctly describe the profile (ZTF19aadgigp: Figure \ref{fig:profile_egs}d), and very boxy profiles (ZTF18achchge: Figure \ref{fig:profile_egs}e) -- all are well described by a circular disk model with typical disk parameters. 

In order to separate the visually classifiable DPEs from other broad-line AGN without a dip between shoulders or a velocity-offset peak, we applied cutoffs to particular disk parameters. We found that requiring an inclination angle $i>14$\textdegree, turbulent broadening $\sigma>600$ km/s, and inner radius $\xi_1<1200$ was extremely effective at identifying almost all classical DPEs with obvious shoulders or asymmetries such as those shown in Figure \ref{fig:profile_egs}, while removing AGN with more symmetric profiles monotonically increasing to the central velocity. The distributions of the three parameters used for DPE classification and their cutoffs are shown in Figure \ref{fig:disksplit}. The cutoffs resulted in automatic classification of 260 sources as DPEs and 1042 as `normal' broad-line AGN. We then visually inspected both samples to ensure that all DPE candidates had either: a clear dip or plateau between shoulders, a velocity offset between the peak of the broad line and the H$\alpha$ narrow line $>500$ km/s, or a $>0.8$ flux ratio asymmetry between red and blue peaks. We found that 22 DPE candidates did not have clear evidence for any of these features and they were moved to the control sample. Similarly, we inspected the control sample for spectra containing the aforementioned features of a DPE profile, and found that 12 DPEs were missed by the disk parameter classification criteria. These were moved from the control sample to the DPE sample. 

\begin{figure*}
\gridline{\fig{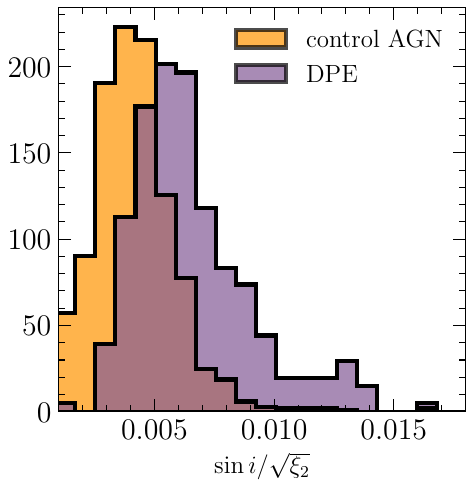}{0.32\textwidth}{} \fig{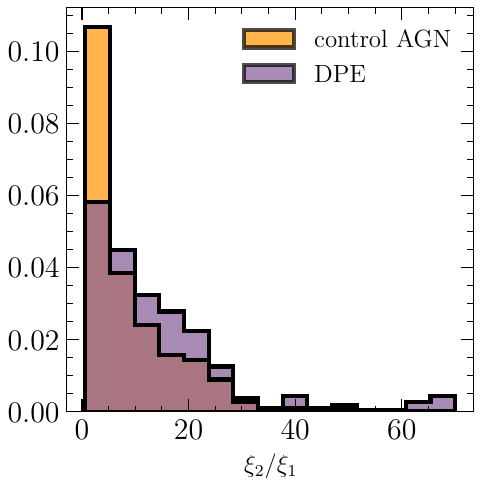}{0.33\textwidth}{} \fig{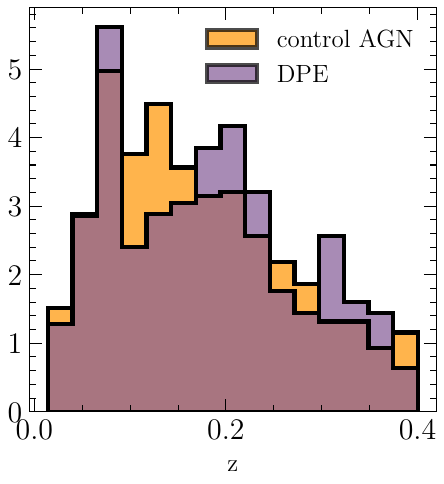}{0.31\textwidth}{}}
\caption{Normalized histograms showing the distributions of three key quantities derived from disk profile fitting after splitting the sample into DPEs and `normal' broad-line AGN. \texttt{Left:} The ratio of the sine of the inclination angle to the square root of the outer radius, which determines the separation between the two peaks. \texttt{Center:} The ratio of the outer to inner radius, which determines the how distinct the two peaks of the profile are and whether the profile is `boxy'. \texttt{Right:} The best-fit redshifts of the DPE sample and the broad-line AGN control sample.}
\label{fig:sampleprop}
\end{figure*}

Our automatic classification procedure, followed by visual reclassification of 34 (3\%) of spectra, resulted in final sample sizes of 250 DPEs and 1052 control AGN. This resulted in 19.2\% of our strongly variable broad-line AGN sample being classified as DPEs. The positions and redshifts of the 250 objects classified as DPEs are presented in Table \ref{table:ztfcands}. The best-fit H$\alpha$ disk parameters of the objects classified as DPEs are shown in Table \ref{table:diskparams} and histograms of each parameter are shown in the Appendix in Figure \ref{fig:diskDPEvsAGN}.

In Figure \ref{fig:sampleprop} we show three key quantities derived from the disk modeling for the DPE and control AGN samples. Firstly, we show that the two samples are well-separated when we plot their distributions of $\sin i/\sqrt{\xi_2}$, which describes the separation between the blue and red shoulders in the circular disk model. This indicates that our sample separation criteria has achieved what we expect - to discriminate between broad-lines with well-separated shoulders, regardless of their relative strengths (which may be affected by phenomena such as spiral arms). Secondly, we show the outer to inner radius ratio $\xi_2/\xi_1$, which determines how distinct the two peaks of the profile appear and how “boxy” the profile appears. We find that the DPEs tend to have larger ratios of $\xi_2/\xi_1$. Finally, we show the redshift distribution of the two populations, where we can see that our classification criteria is not strongly biased by redshift for our redshift range of $z<0.4$.  

\section{Variability analysis}

\subsection{Construction of optical and mid-IR light curves}

In order to produce light curves of the DPEs and AGN control sample using both positive and negative photometry from ZTF difference imaging, we used the ZTF forced photometry service \citep{Masci2019TheArchive}. We extracted all available photometry from the ZTF public and partnership fields between 2018-01-01 and 2023-05-01. After removing poor quality images by requiring the \texttt{procstatus} flag be $=0$, we measured the baseline flux from  the reference images, applied zeropoints, and combined the baseline flux measured from the reference images and the single epoch fluxes to produce g- and r-band light curves of the two samples. 

\begin{figure*}
\gridline{\fig{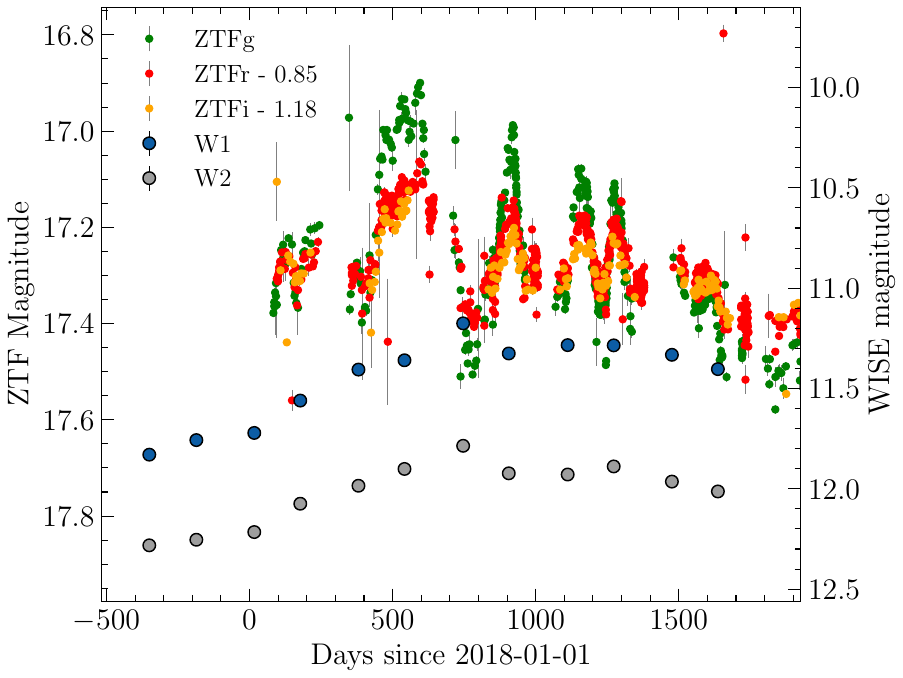}{0.48\textwidth}{a) ZTF18aarippg} 
 \fig{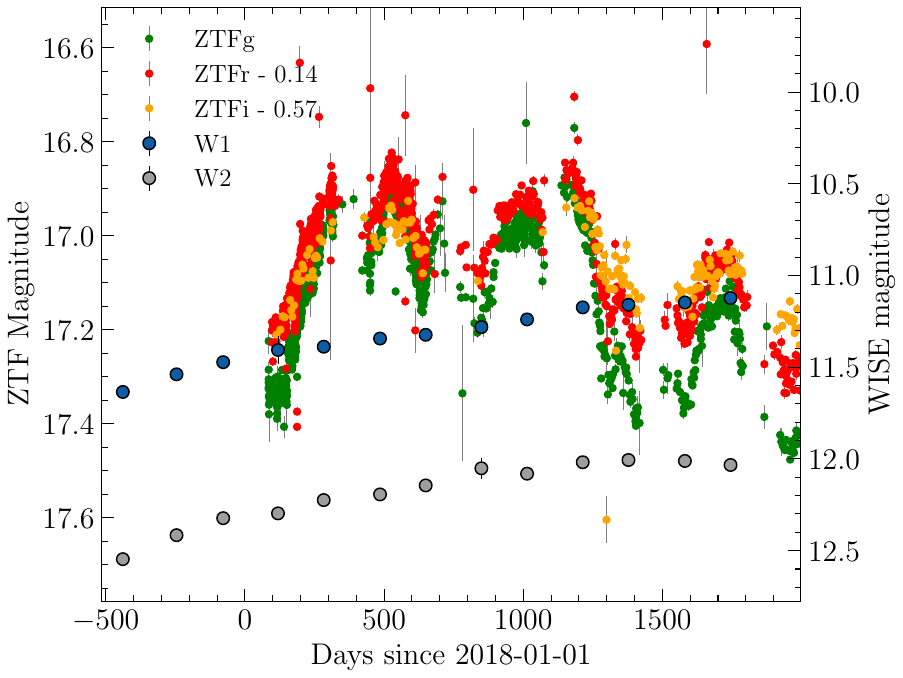}{0.48\textwidth}{b) ZTF18aaymybb}} 
 \gridline{
\fig{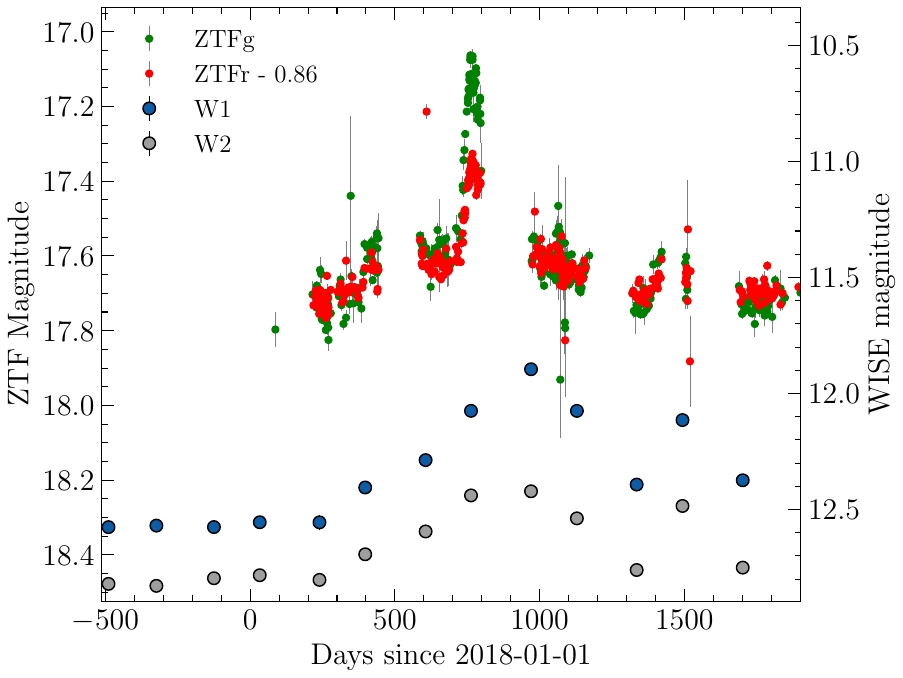}{0.48\textwidth}{c) ZTF19aagwzod} \fig{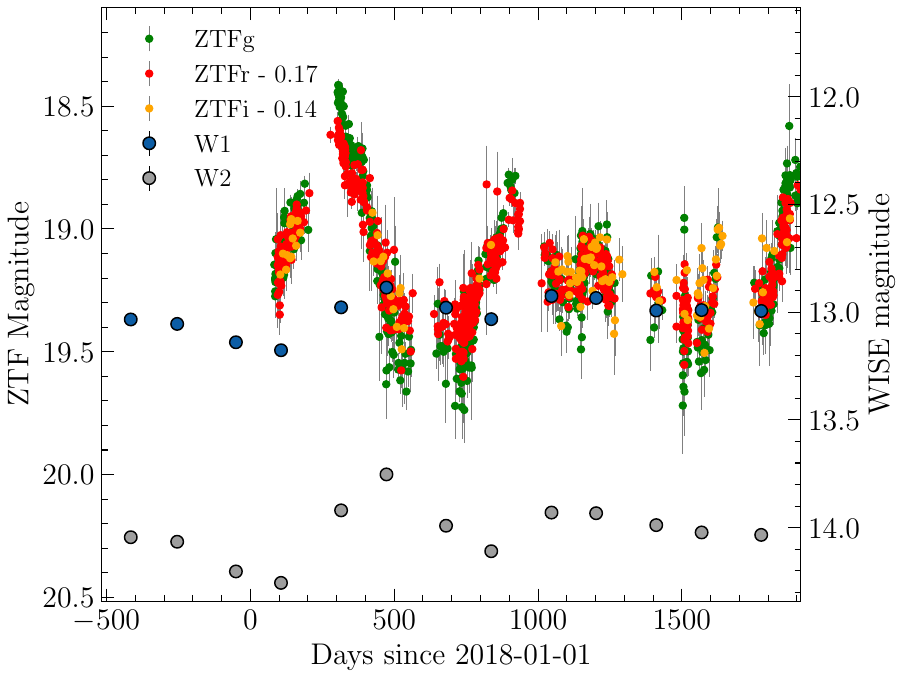}{0.48\textwidth}{d) ZTF18aaznjgn}}
\gridline{ \fig{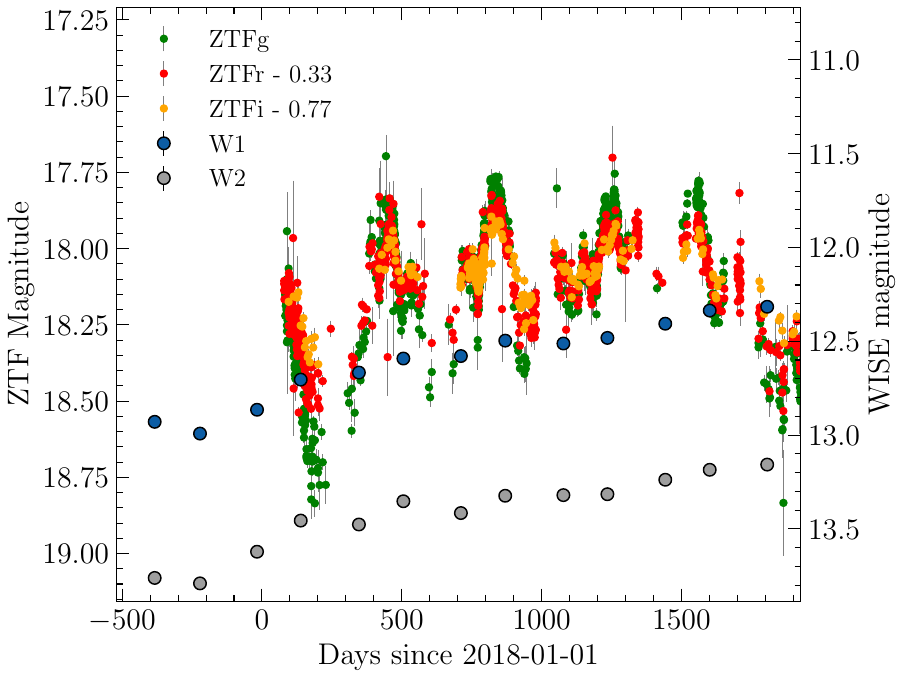}{0.48\textwidth}{e) ZTF18aalslhk} \fig{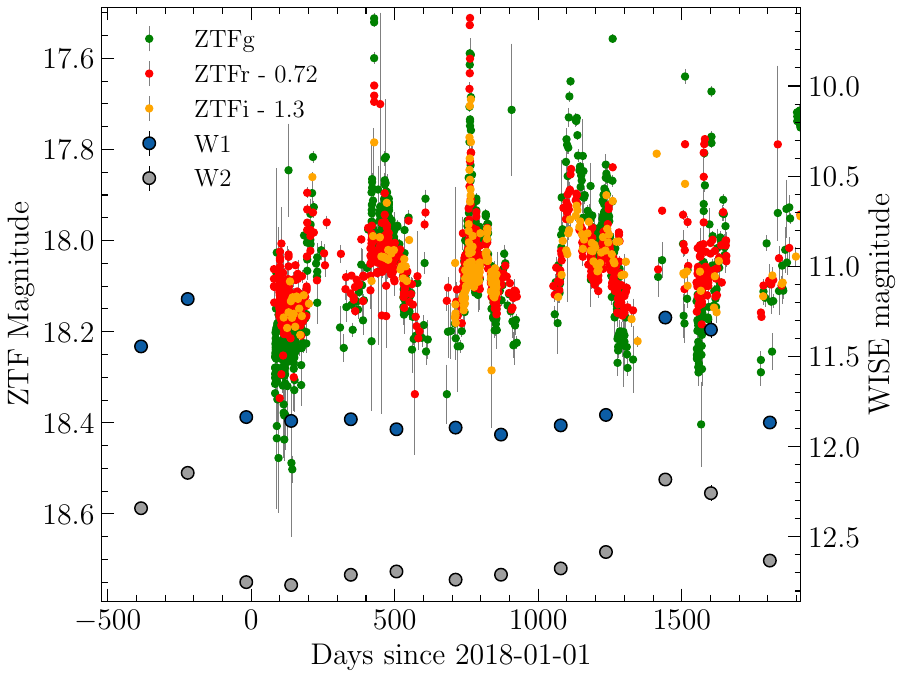}{0.48\textwidth}{f) ZTF18aakehue}}
\caption{Notable optical (ZTF) and mid-IR (\textit{WISE}) light curves of selected DPEs. The left y-axes display the ZTF AB magnitudes while the right y-axes display the \textit{WISE} Vega magnitudes. We include light curves of previously reported SMBH merger candidate (ZTF18aarippg (a)), a DPE with a complex and time-varying double-peaked profile shown later in Figure \ref{fig:monitor} (ZTF18aaymybb (b)), a previously reported CLAGN candidate (ZTF19aagwzod (c)), a DPE with a clear IR echo (ZTF18aaznjgn (d)), and two DPEs exhibiting quasi-regular fluctuations (ZTF18aalslhk (e) and ZTF18aakehue (f)).}
\label{fig:lcs_notable}
\end{figure*}

Examples of optical light curves of selected DPEs are shown in Figure \ref{fig:lcs_notable}. We present the updated ZTF light curve of ZTF18aarippg, the previously reported candidate for an inspiraling SMBH binary \citep{Jiang2022Tick-Tock:Binary}. We have also selected the particular examples ZTF18aalslhk and ZTF18aakehue because, by eye, it appears that the light curves may be better fit by a damped harmonic oscillator model over a typical AGN damped random walk model, making them most comparable to ZTF18aarippg. We note that apparently quasi-periodic variability arises naturally in a fraction damped random walk light curves, but we nonetheless present these specific cases as possible subjects of interest for periodicity analysis in future work. We also present the light curves of 2 DPEs for which we present time-domain spectroscopic monitoring in Section 5 (ZTF18aaymybb, and ZTF19aagwzod).

The AGN and DPE samples also had recent mid-IR photometry available in W1 (3.4$\mu$m) and W2 (4.6$\mu$m) bands from the \textit{WISE} mission \citep{Mainzer2011NEOWISERESULTS, Mainzer2014InitialMission}. We obtained the neoWISE light curves from IRSA \citep{NEOWISETeam2020NEOWISE-RTable}. NeoWISE observes each field with a $\sim6$ month cadence, taking multiple observations over a short $<2$ day period. We report the median and standard deviation of the observations taken upon each $\sim6$ monthly visit to the field. The mid-IR light curves of selected DPEs are also shown in Figure \ref{fig:lcs_notable}. 

In a number of cases, the WISE light curves show the presence of mid-IR dust echoes with delays of $\sim200$ days relative to the optical (e.g. ZTF19aagwzod, ZTF18aaznjgn, as well as ZTF18acvcadu which is not pictured). In other cases the delays appear to be very long, on timescales of $>1000$ days (ZTF18abzweee, not pictured, had such a delay). For a large fraction of cases, the WISE light curve follows the long term variation of the optical light curve, but we do not resolve shorter timescale ($<$1\ year) variability (e.g. ZTF18aalslhk, ZTF18aaymybb). Future work could investigate the relationship between best-fit inclination angles from the disk model fitting and the delay of the mid-IR echo from the dusty torus. The ZTF and WISE light curves of the full AGN sample have also been made available in the github repo containing the intermediate data products (see footnote 1).

\subsection{Power spectrum analysis}

In order to quantify the characteristic timescales and amplitude of optical variability in the DPE and control AGN populations, we generated power spectra of the ZTF light curves. We adopted the following method to produce g and r-band power spectra from the unevenly sampled ZTF data. To prepare the light curves for power spectra production we first removed low-significance observations with uncertainties $>$10 times the median uncertainty.  To reduce outliers we normalized the fluxes by the best-fit linear trend and removed data which deviated by more than 7 median absolute deviations. This outlier-removal approach may result in smoothing or suppression of rapid fluctuations in the light curve, but is nonetheless required to remove poor quality photometric data points. 

\begin{figure*}
\gridline{\fig{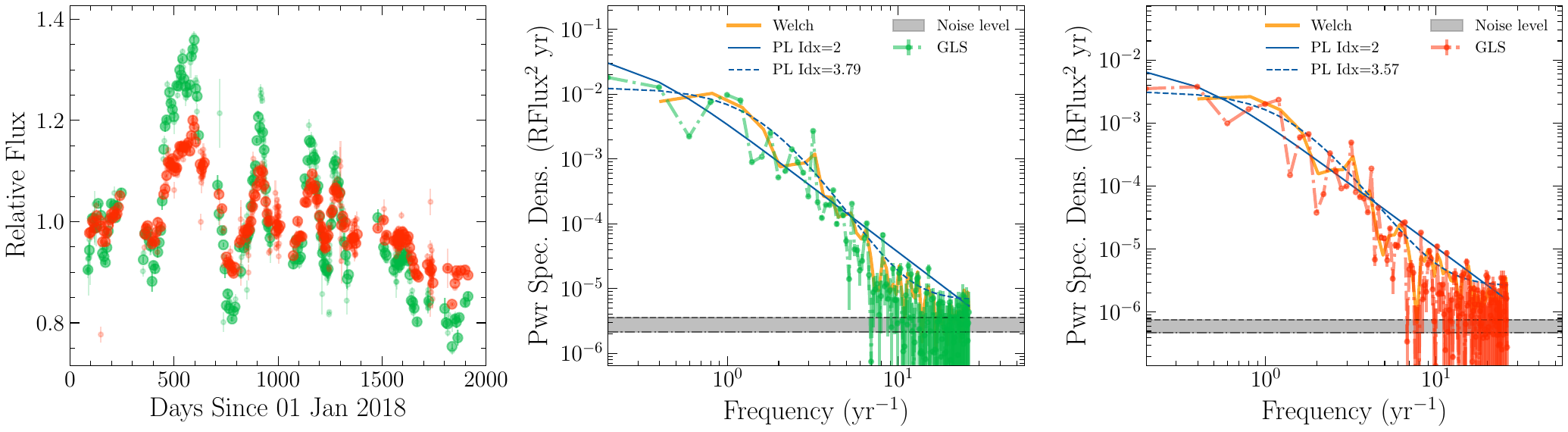}{0.87\textwidth}{ZTF18aarippg}}
\gridline{\fig{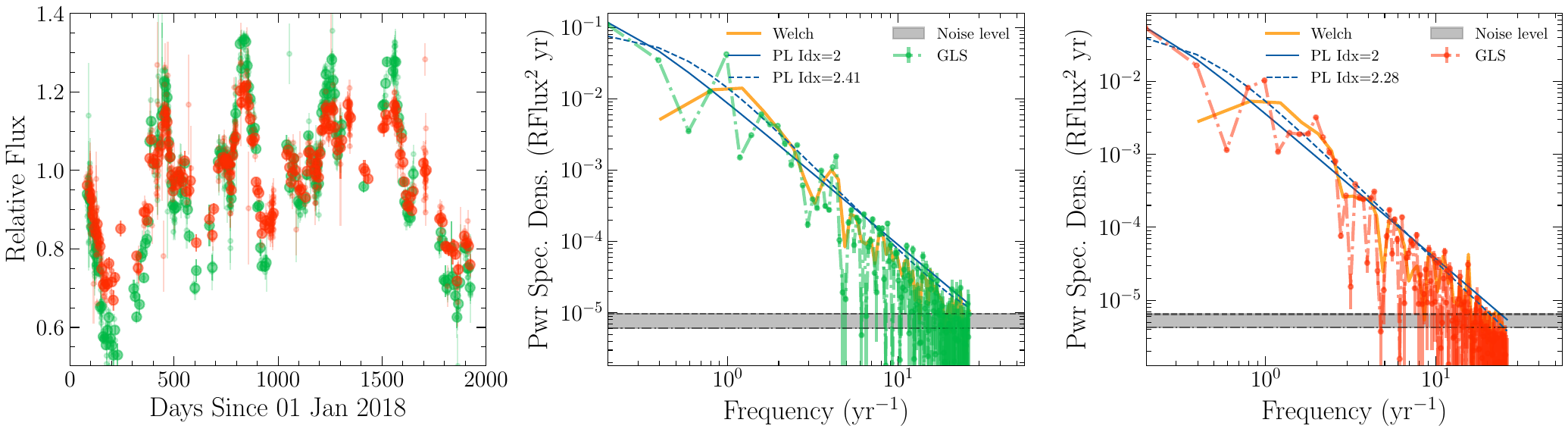}{0.87\textwidth}{ZTF18aalslhk}}
\gridline{\fig{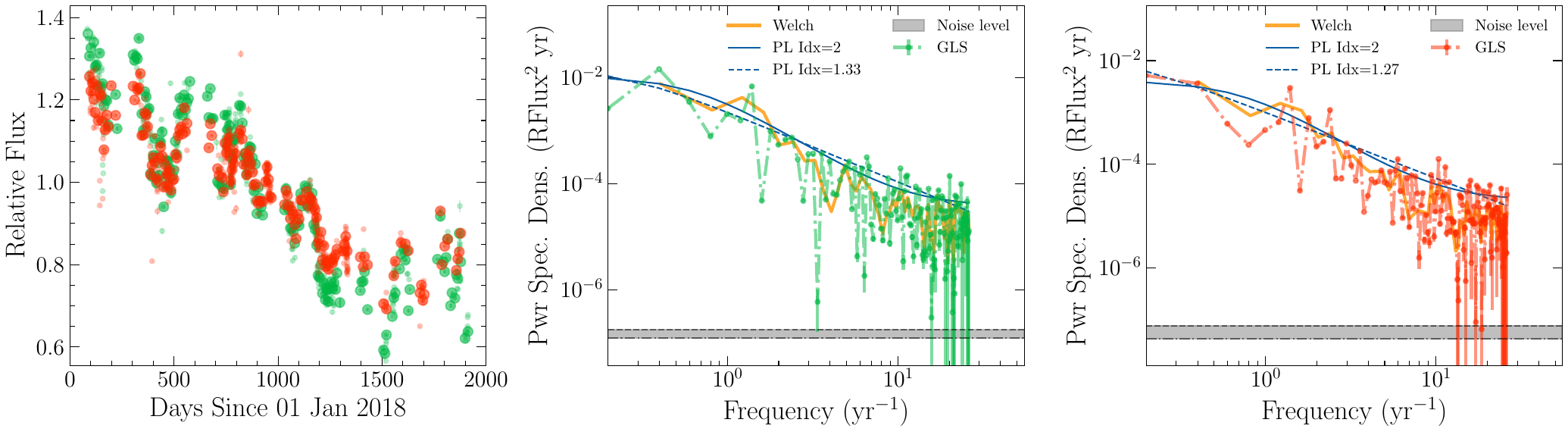}{0.87\textwidth}{ZTF18aaadgxi}}
\gridline{\fig{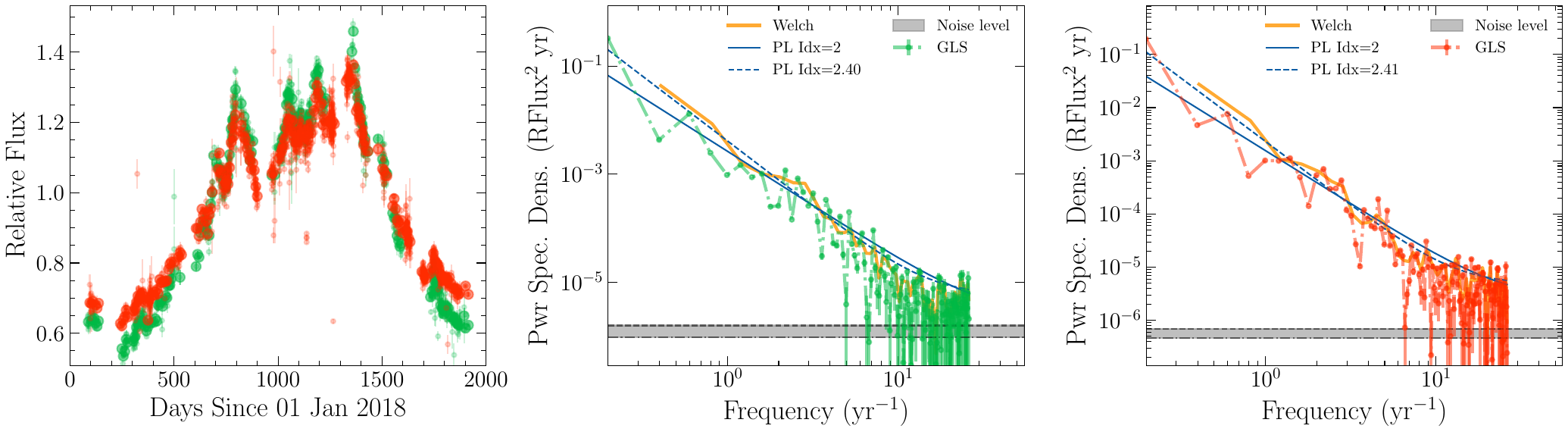}{0.87\textwidth}{ZTF18aacjtlo}}
\caption{For three DPE and one control AGN example \texttt{Left}: The g- and r-band relative flux vs time from ZTF;  \texttt{Center:} The power spectrum of the g-band light curve derived from the GLS method (green), the Welch periodogram (orange), the best-fit power law + white noise + low frequency turnover model with free power law index (blue solid), the best-fit power law + white noise + low frequency turnover model with power law index fixed to 2 (blue dashed), and the light curve noise range estimate (gray shaded); \texttt{Right:} The same as above but for the r-band light curve. These four power spectra were selected to display the range of power law spectral indices and high frequency turnovers observed in the sample.}
\label{fig:powerspec}
\end{figure*}

\begin{figure*}
\gridline{\fig{figures/psdanalysis/AMpdist_DPE.pdf}{0.32\textwidth}{}\fig{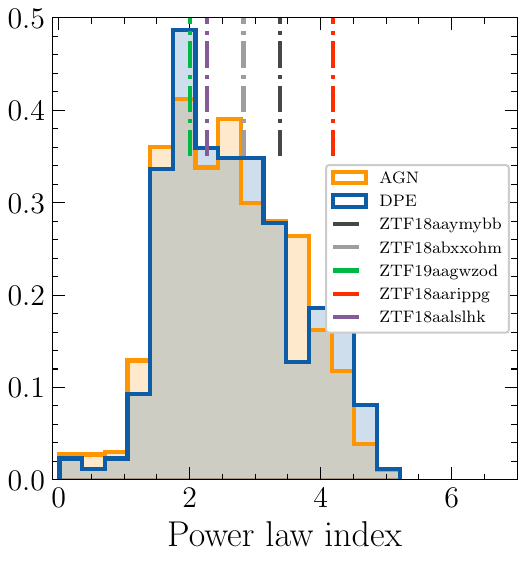}{0.32\textwidth}{}\fig{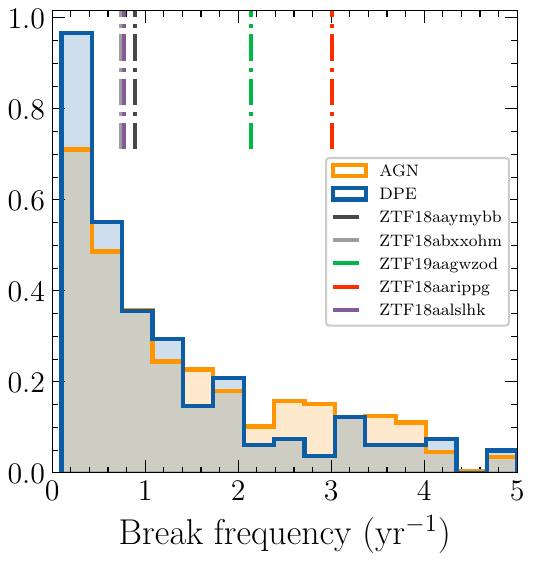}{0.32\textwidth}{}}
\caption{Normalized histograms of three parameters derived from fitting of g-band ZTF power spectra: PSD amplitude, power law index and position of the low frequency break. We show the distributions for the DPE and control AGN samples, and note the locations of parameters derived for 5 notable DPEs: the SMBH merger candidate (ZTF18aarippg), the flaring CLAGN candidate (ZTF19aagwzod), a DPE with a quasi-periodic signal (ZTF18aalslhk), and two DPEs with unusual double-peaked profiles (ZTF18aaymybb and ZTF18abxxohm; see section 5). Note that break frequencies less than 1.8 (the inverse of $0.1\times$ the light curve baseline) may be unreliable.}
\label{fig:psd_hists}
\end{figure*}

We next binned the data to uniform time bins with full width 7.0\,d.  For each bin we estimated the mean flux as the uncertainty-weighted sum of the individual fluxes.  The binned measurements were more robust against single-observation outliers, so we further filtered out flux points which differed from the resulting mean values by $>$5$\sigma$, where $\sigma$ refers to the uncertainty on the average flux obtained by propagation of uncertainty.  We did this iteratively until no individual outliers remained.  We next eliminated bins with large flux uncertainties, $>10$ times the median, typically those which contain only a single low-significance observation.  To identify outlying time bins, we computed the difference in mean flux between each bin and its neighbors, computed the standard deviation of this population of differences, and removed any time bins with a difference greater than 5 standard deviations.  The resulting uniformly binned light curves generally retained $>$90\% of the original data, and most obvious outliers were automatically removed.

We computed the power spectral density with a generalized least squares (GLS) method following the approach of \citet{Coles2011PulsarNoise}.  Specifically, we adopted a model for the data in the time domain, 

\bigskip\noindent
$$
F(t) = F_0 + F_1 (t-t_0) + 
\sum_{k=1}^{N} 
\Bigg\{ a_k \sin \left[2\pi k (t-t_0)\over T\right] 
\quad
$$
\begin{equation}
\hbox to 10em{\hss}
+ b_k \cos \left[2\pi k(t-t_0)\over T\right]\Bigg\},     
\end{equation}
with $T$ the total data span. In other words, the model comprises a mean flux, a linear flux ramp, and a Fourier series with coefficients $a_k$ and $b_k$ describing the variability.  To constrain the many degrees of freedom in the Fourier series, we assumed a model for the power spectral density (PSD, P(f,\ $\lambda$)) and that the Fourier coefficients were distributed as a normal distribution with width $\sqrt{P}/2$. Our final model for P(f) was a power law of index $\gamma$ with an additional white noise component (W) at high frequencies and a turnover to a flat spectrum at low frequencies ($f_c$):
\begin{equation}
P(f) = W+A \left(f^2 + f_c^2\over 1 + f_c^2\right)^{-\gamma/2}
\end{equation}

To determine the parameters of these models, $\lambda$, along with the time-domain components $F_0$, $F_1$, and the Fourier coefficients, we used generalized least squares optimization. We selected $N$, the number of Fourier components, as half of the number of data points, such that the highest frequency $N/T$ was the Nyquist frequency. The resulting fit simultaneously provided estimates of the PSD parameters, the PSD itself (via the Fourier components), and the total log likelihood for the model. We then used emcee \citep{Foreman-Mackey2013ttemcee/ttHammer} to sample over each free parameter for 500 iterations with a burn-in time of 100 iterations to determine the best-fit values and $1\sigma$ uncertainties for the amplitude, power spectral index, white noise level and low frequency turnover. We also produced an alternative set of models with a power law index fixed to a value of 2, for comparison to the models in which the power law index was allowed to vary. Examples of 4 DPE light curves and their corresponding power spectra are shown in Figure \ref{fig:powerspec}.

For comparison to the power spectral densities (PSDs) generated via the GLS method, we additionally estimated a model independent power spectral density using Welch's method, based on the weighted, overlapped sum of Hann-windowed Fourier transforms.  Our choice of 128 data points per segment reduced noise while retaining reasonable sensitivity to low-frequency power.  The PSD estimate obtained in this way is shown in orange solid lines in the power spectra plots of Figure \ref{fig:powerspec}.

The power spectra shown in Figure \ref{fig:powerspec} demonstrate how a high frequency turnover is required in a fraction of cases in order to model an intrinsic white noise component above the noise level which naturally arises from flux uncertainties in the light curve (shown in gray). ZTF18aarippg is an example of a DPE with a clear high frequency white noise component. In other cases, (e.g. ZTF18aalslhk) the power law reaches the light curve noise level before the intrinsic AGN white noise induces a high frequency turnover. The spectral indices of the PSDs have a large range: some objects, such as ZTF18aacjtlo, have steep spectral indices of $\sim2.4$, while other objects, such as ZTF18aaadgxi, have very shallow power spectra with spectral indices $\sim1.3$.

\subsection{Variability properties of the DPE and control AGN samples}

The best-fit power spectrum parameters derived from ZTF g-band light curves are shown for all DPEs in Table \ref{table:ztfcands}. The distributions of power spectral index, PSD amplitude and turnover frequency are shown in Figure \ref{fig:psd_hists}. The DPE sample had a median log amplitude and standard deviation of $-2.65$ and $0.49$ respectively, and a median power law index and standard deviation of $2.50$ and $0.96$. To find the median break frequency we first removed values $<1.8$ yr$^{-1}$ (the inverse of $0.1\times$ the light curve baseline, where the best-fit turnovers may be unreliable \citep{Burke2021ADisks}) as well as outliers $>6$ yr$^{-1}$ arising from poor fits. We found a median break frequency and standard deviation of $0.77$ and $0.7$ yr$^{-1}$ respectively for the DPE sample. By comparison, the control AGN population had a median log amplitude and standard deviation of $-2.68$ and $0.48$, a median power law index and standard deviation of $2.55$ and $0.94$, and a median break frequency and standard deviation of $1.1$ and $0.7$ yr$^{-1}$. Because the high frequency white noise turnover was only detected at sufficient S/N for a fraction of light curves, we do not report the white noise levels for each population.

We applied a two sample KS test to each power spectrum parameter to determine the probability with which we can reject the null hypothesis that the AGN and DPE power spectrum parameters were drawn from the same distributions. For the amplitude and power law index parameters we obtain p-values of 0.46 and 0.52, indicating that we do not have evidence that they were drawn from different distributions. For the turnover frequency parameter we obtain a p-value of 0.0072, so we can reject the hypothesis that they were drawn from the same distribution at $>3\sigma$. We also note that a power law index $>2$ was ruled out to 95\% confidence for only 10.1\% of AGN and 11.1\% of DPEs.

We searched for correlations between the PSD amplitude, turnover frequency and spectral index parameters from the power spectra and each disk geometry parameter derived from the spectroscopic fits of the DPE sample. We calculated the Spearman correlation coefficient and associated p-value for each parameter combination and found no evidence for correlations between optical variability parameters and accretion disk geometry parameters.

In summary, our analysis of the power spectra of optical ZTF light curves finds a wide distribution of variability amplitudes, power law spectra index, and low frequency turnovers for the ZTF AGN population. We do not find significant differences between the variability amplitudes and power law spectral indices of the DPE and control AGN populations, but we do find some evidence that the low frequency break occurs at lower frequencies for DPEs compared to `normal' broad-line AGN. 

\section{Radio detections and jet imaging}

We undertook a search for radio emission from the DPE and control AGN samples in the Karl G. Jansky Very Large Array Sky Survey \citep[VLASS;][]{Lacy2020TheDesign}. This survey covers a total of 33,885 deg$^2$ in the 2-4 GHz range with an angular resolution of $\sim 2^{\prime\prime}.5$ and will obtain a coadd $1\sigma$ sensitivity of 1 $\mu$Jy/beam by survey end in 2024. We searched for crossmatches within $10^{\prime\prime}$ in Table 2 of the VLASS Epoch 1 and 2 Quick Look Catalogues which contains $\sim 700,000$ compact radio sources with $>1$ mJy/beam detections associated with mid-IR hosts from the un\textit{WISE} catalog \citep{Gordon2021ASurvey}.

\begin{figure*}
\gridline{\fig{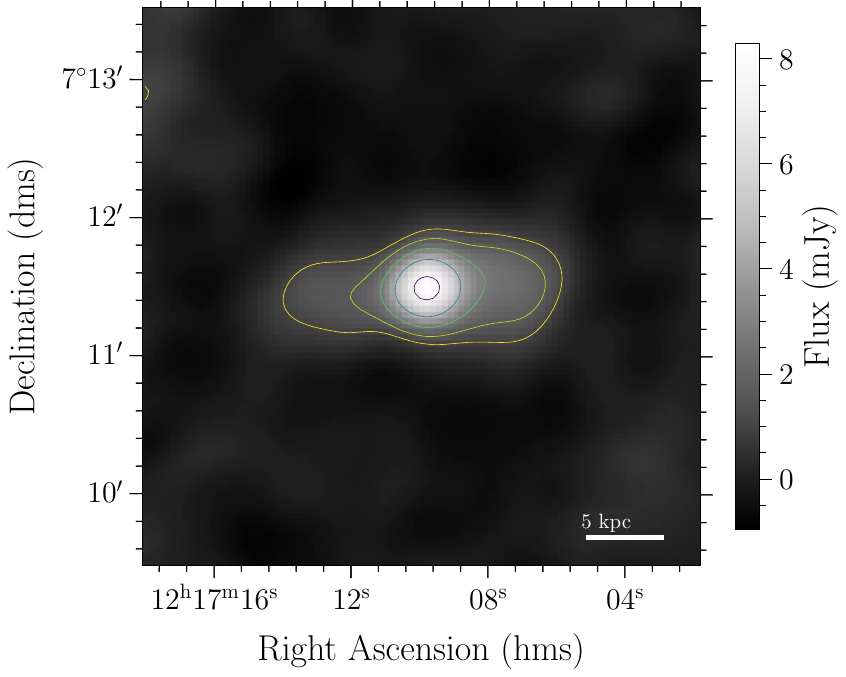}{0.45\textwidth}{ZTF18aarywbt (RACS)} 
 \fig{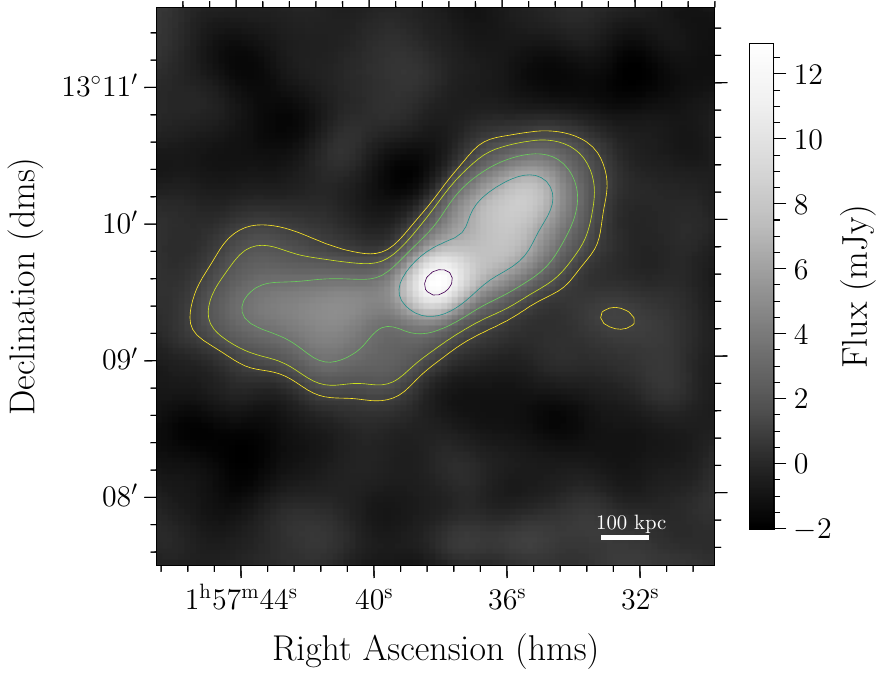}{0.45\textwidth}{ZTF19abizomu (RACS)}} 
 \gridline{\fig{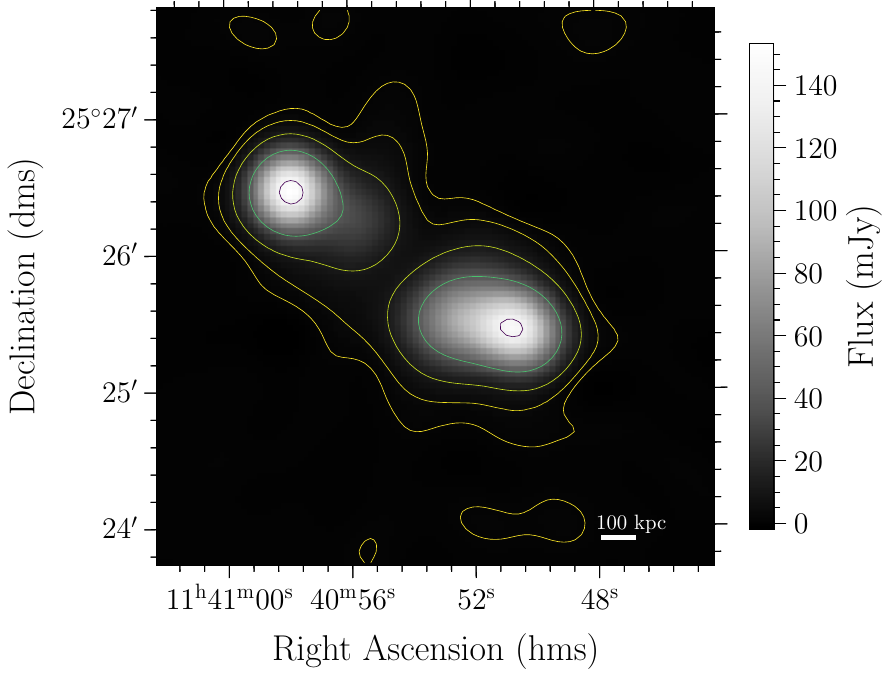}{0.45\textwidth}{ZTF18aaqdmih (RACS)}
\fig{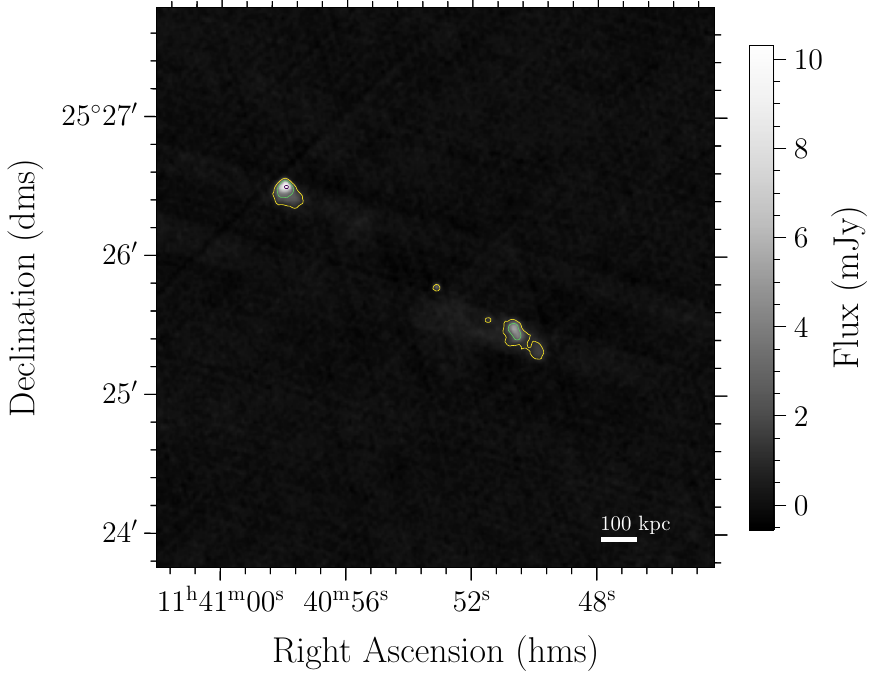}{0.45\textwidth}{ZTF18aaqdmih (VLASS Epoch 1.2)}}
\caption{Radio imaging cutouts from RACS and VLASS showing the radio jets around three DPEs with intermediate inclination angles: ZTF18aarywbt ($i=19^{+1}_{-1}$), ZTF19abizomu ($i=14^{+1}_{-2}$) and ZTF18aaqdmih ($i=35^{+2}_{-2}$). The grayscale colors indicate the observed fluxes in linear space while the contour intervals are in log space. In all cases the AGN is in the image center.}
\label{fig:radioimages}
\end{figure*}

We also searched for radio emission in the Rapid ASKAP Continuum Survey (RACS), with first epoch observations covering the whole southern sky to +41 deg declination with the Australia Square Kilometre Array Pathfinder at a central wavelength of 887.5 MHz \citep{Hale2021TheRelease}. We crossmatched our sample with a $10^{\prime\prime}$ radius to the first Stokes I Source Catalogue Data Release, which has an estimated 95\% point source completeness at an integrated flux density of $\sim3$ mJy.

The radio fluxes and non-detections from the two surveys are displayed in Table \ref{table:ztfcands}. Of the DPEs, 29 of 235 in the VLASS survey area were detected (12.3\%) while 23 of 121 in the RACS survey area were detected (19.0\%). 66 of 1239 control AGN within the VLASS survey area (5.3\%) were detected and 104 of 576 control AGN in the RACS survey area (18.1\%) were detected.  DPEs were therefore 2.3 times more likely than the control AGN to be detected at 20cm in VLASS. By contrast, DPEs were just as likely as the control AGN to be detected at 34cm in RACS. 

For those objects with radio detections, we compared the distributions of radio fluxes measured from VLASS imaging between the DPE and control AGN samples and found that they were very similar. A two-sample Kolmogorov-Smirnov test found no evidence to reject the null hypothesis that the radio fluxes of the two samples were drawn from the same distribution with a p-value of 0.24. We also found no evidence that any of the disk parameters derived from the spectroscopic modeling were significantly different between the radio-detected and radio-undetected samples.

We compared the radio fluxes across the two available VLASS epochs to determine the fraction of DPEs and control AGN with $>3\sigma$ variability between 2017-2018 and 2018-2021. We found that 6\% of DPEs and 25\% of the control AGN with radio detections exhibited flux changes greater than 3 times the flux uncertainties between the two epochs. 

We checked the RACs and VLASS imaging of radio-detected DPEs using the CIRADA Image Cutout Web Service\footnote{http://cutouts.cirada.ca/} to determine if the radio detections associated with DPEs were all point sources or if there were additional radio jets visible in the imaging. We noted that 3 DPEs (ZTF18aarywbt, ZTF19abizomu and ZTF18aaqdmih, which had inclination angles of $14<i<35$), had large radio jets emanating from a point source coincident with the AGN. The RACS and VLASS imaging showing these jet structures is shown in Figure \ref{fig:radioimages}. 

\section{Spectroscopic evolution of DPEs}

\begin{figure*}
\gridline{\fig{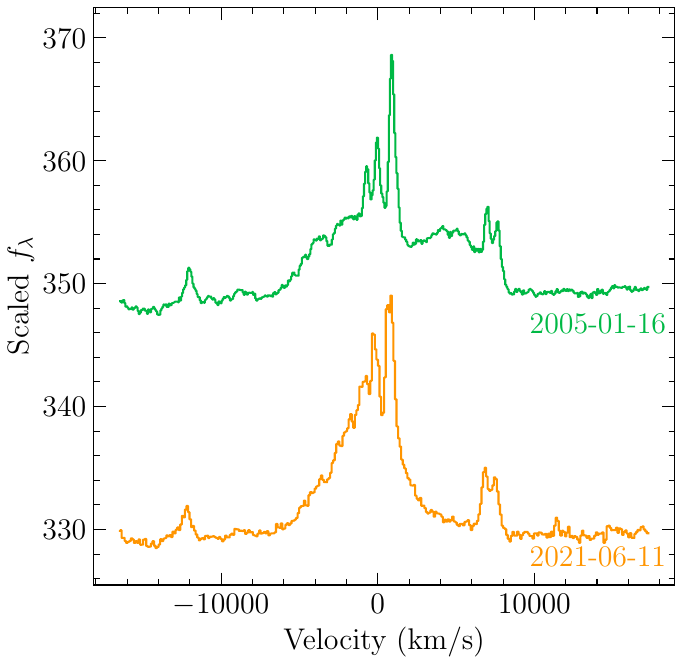}{0.25\textwidth}{a) ZTF18aarywbt}\fig{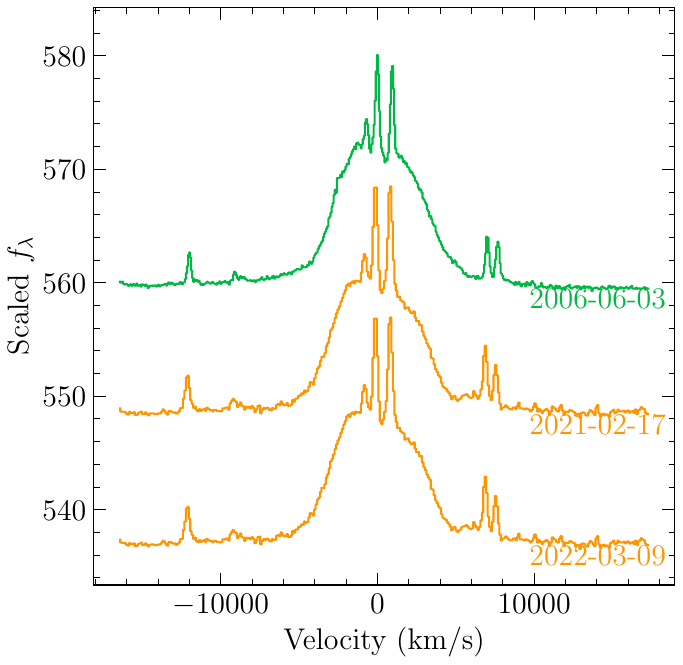}{0.25\textwidth}{b) ZTF18aaylbyr}\fig{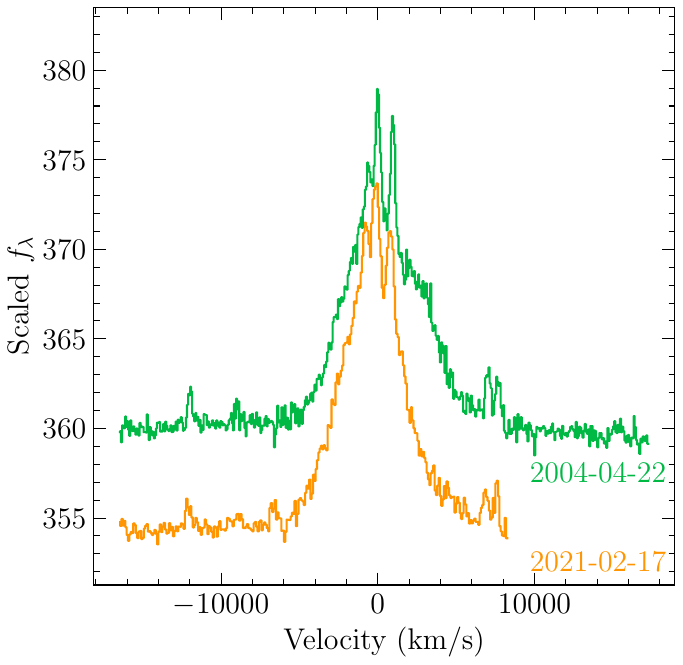}{0.25\textwidth}{c) ZTF18aalslhk}\fig{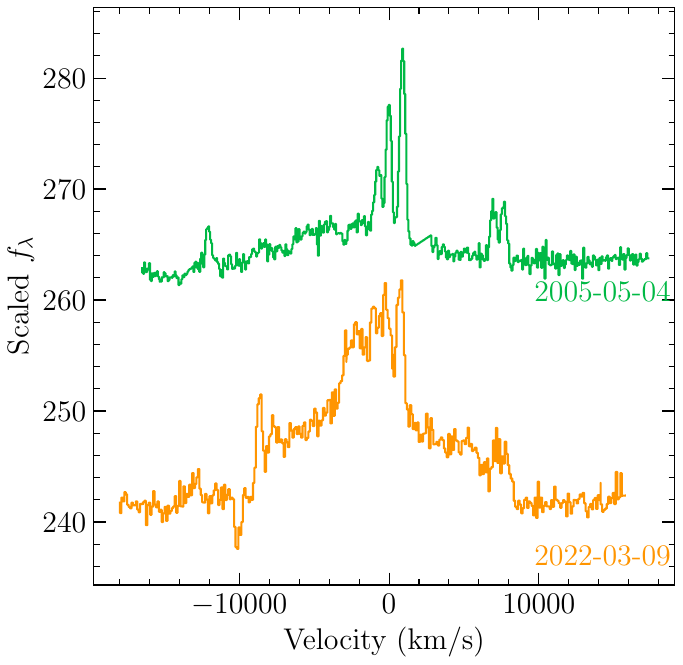}{0.25\textwidth}{d) ZTF18aarippg}}
\gridline{\fig{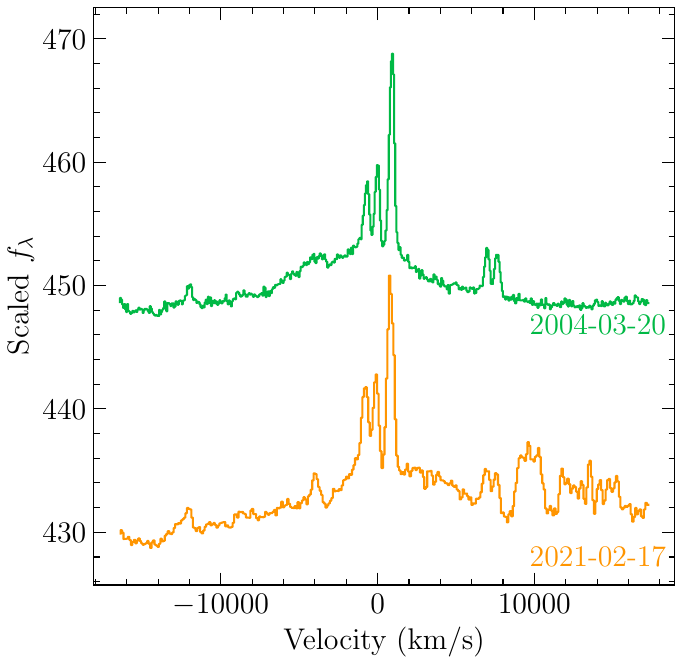}{0.25\textwidth}{e) ZTF19aarlffl}\fig{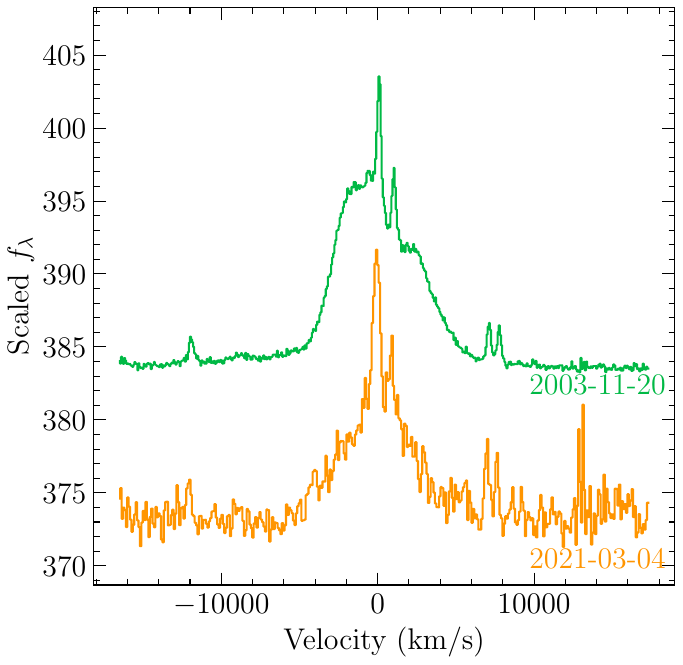}{0.25\textwidth}{f) ZTF19acbvtcx}\fig{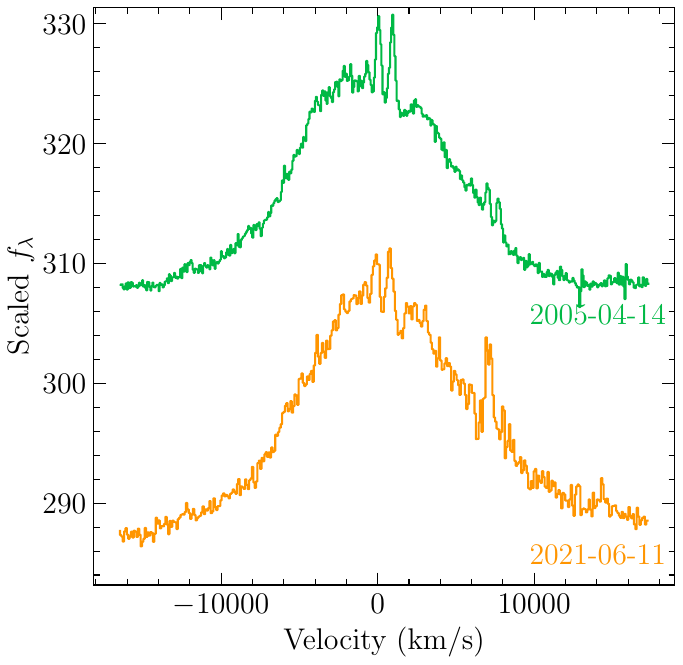}{0.25\textwidth}{g) ZTF18acqtdnj}\fig{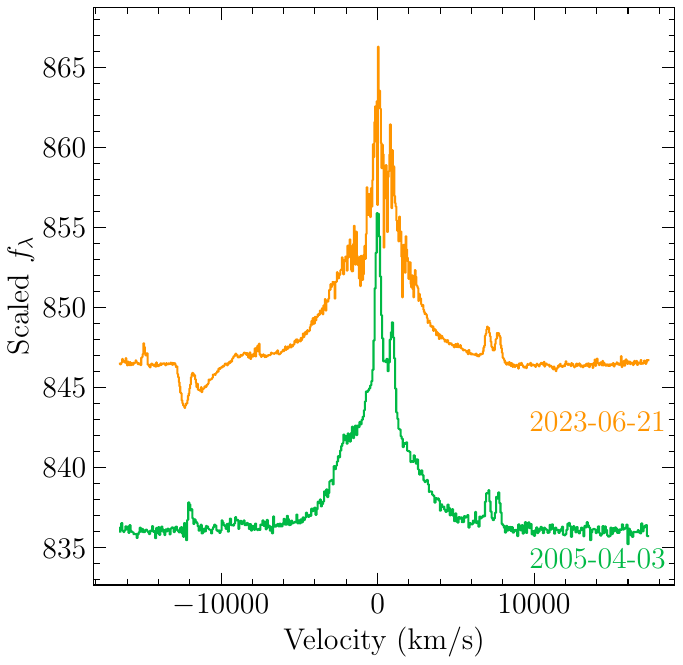}{0.25\textwidth}{h) ZTF19aautrth}}
\gridline{\fig{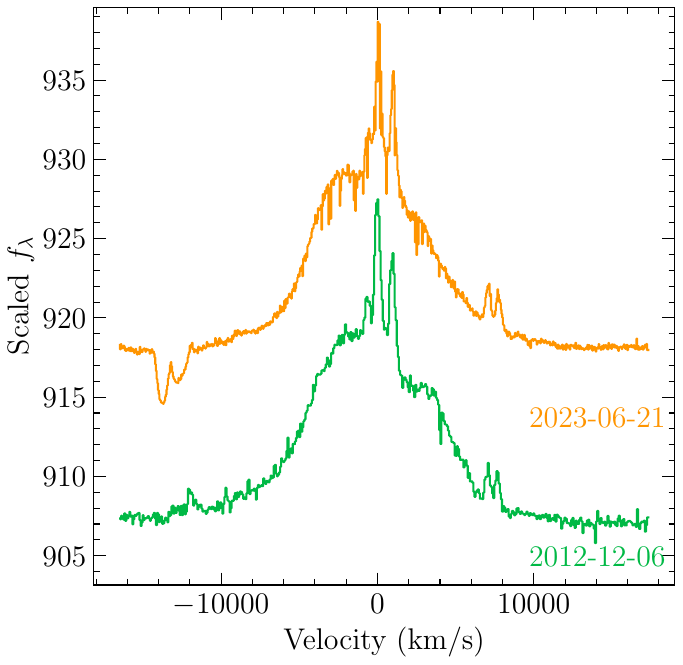}{0.25\textwidth}{i) ZTF19aayrjsx}\fig{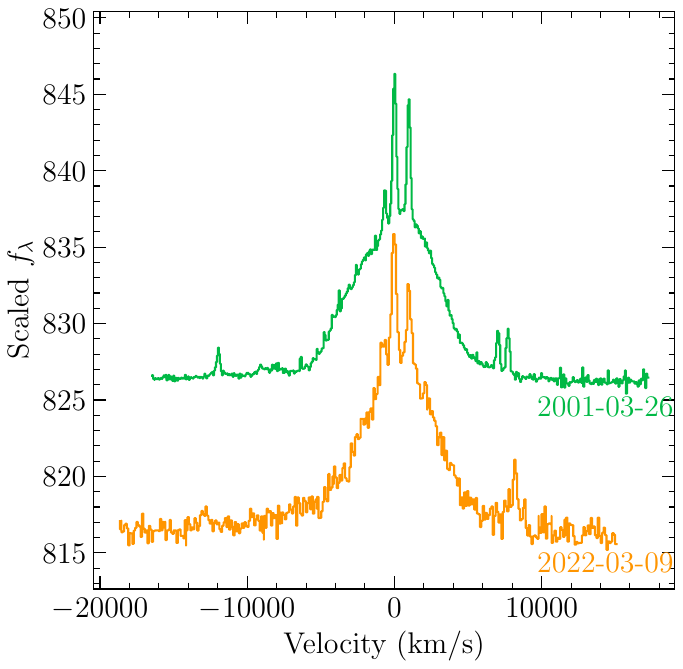}{0.25\textwidth}{ j) ZTF18ablqgje}\fig{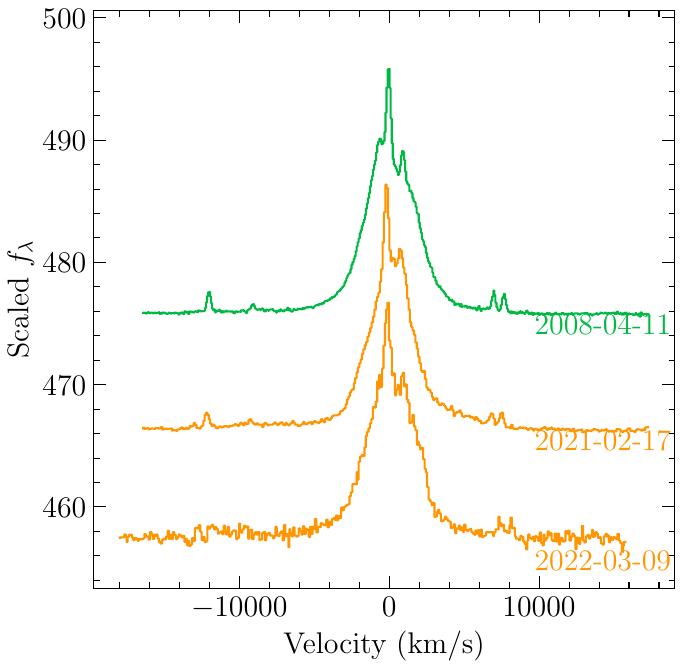}{0.25\textwidth}{k) ZTF18aatxsvu}\fig{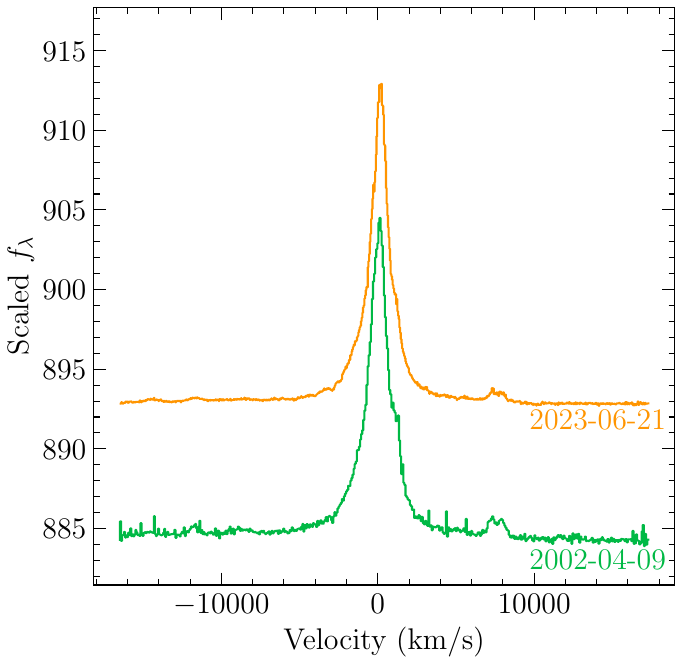}{0.25\textwidth}{l) ZTF19aadggaf}}
\caption{Comparison of Balmer broad line structures from recent spectra LDT or Keck spectra (orange) and archival SDSS spectra (green) of 12 DPEs. Six of twelve show changes in the relative fluxes of the blue and red shoulder. }
\label{fig:longmonitor}
\end{figure*}

In order to search for changes in disk morphology over 10-20 year timescales we obtained follow-up spectra of 12 DPEs for comparison to archival SDSS spectra. Spectra were taken with either the DeVeny spectrograph on the Lowell discovery Telescope using a $1.5^{\prime\prime}$ slit, central wavelength of 5700 Å, a spectroscopic coverage of 3600-8000 Å and total exposure times ranging from 1000-3000 s. Three spectra were taken with the Keck LRIS spectrograph using a $1.0^{\prime\prime}$ slit, a 400/8500 grating, and 600/4000 grism to obtain spectroscopic coverage over 3500-9500\AA, and total exposure times of 1500 s. Comparisons between recent LDT/LRIS and archival SDSS spectra, with time intervals ranging from 13-18 years, are shown in Figure \ref{fig:longmonitor}. 

Of the 12 objects, six show notable changes in the relative fluxes or positions of the two shoulders in the double-peaked profile. ZTF19aarlffl (Figure \ref{fig:longmonitor}e), which had a bright blue shoulder in 2004, exhibited instead a prominent red shoulder in 2021. ZTF18aarywbt (Figure \ref{fig:longmonitor}a) and ZTF18aalslhk (Figure \ref{fig:longmonitor}c), which had bright blue and red shoulders in 2005 and 2004 respectively, now exhibit only blue shoulders with a smoother shape. ZTF19aayjrsx (Figure \ref{fig:longmonitor}i), an off-nuclear AGN candidate from \citet{Ward2021AGNsFacility} showed a decrease in the peak velocity of the red shoulder, while ZTF19aautrth (Figure \ref{fig:longmonitor}h), another off-nuclear AGN candidate, shows no line profile changes. ZTF18aarippg (Figure \ref{fig:longmonitor}d) now has prominent blue and red shoulders and high velocity when it previously only had a fainter blue shoulder, as previously noted by \citet{Jiang2022Tick-Tock:Binary}. Such substantial variations in relative flux of the blue and red shoulders have been noted in many other DPEs such as Arp 102B, 3C 390.3, NGC 1097, NGC 7213, 3C 59 and 1E 0450.3–1817 and have been modeled by precession of hotspots and spiral arms \citep{Storchi-Bergmann2002Double-peaked1097,Sergeev2002Variability2000,Gezari2007LongTermNuclei,Lewis2010Long-termLines,Jovanovic2010Variability390.3,Popovic2011Spectral19952007,Schimoia2012ShortNGC1097,Shapovalova2013Spectral102B,Popovic2014,Schimoia2017EvolutionNGC7213}. 

\subsection{Spectroscopic monitoring of two unusual DPEs and a candidate CLAGN}

While undertaking spectroscopic follow-up of ZTF transients, we serendipitously discovered two new and atypical DPEs. ZTF18aaymybb was observed to have a complex double-peaked profile with a large dip between the central wavelength and the red shoulder which cannot be well-described by a circular disk model. Another object, ZTF18abxxohm, had two triangular-shaped broad lines: one at the rest H$\alpha$ wavelength and one at a velocity of $\sim2500$ km/s from the rest wavelength. We also undertook detailed spectroscopic follow-up of changing-look AGN candidate, ZTF19aagwzod, which was discovered after an optical flare to have transitioned from a Seyfert 1.9 to Seyfert 1 classification \citep{Frederick2020AX-rays}. This object has since been the subject of detailed multi-wavelength follow-up indicating the presence of X-ray variability but no X-ray spectral evolution \citep{Saha2023MultiwavelengthSeyfert}. The broad-line profile of ZTF19aagwzod is typical for DPEs which are well-described by a circular disk model. The disk parameters can be found in Table \ref{table:diskparams}. 

\begin{figure*}
\gridline{\fig{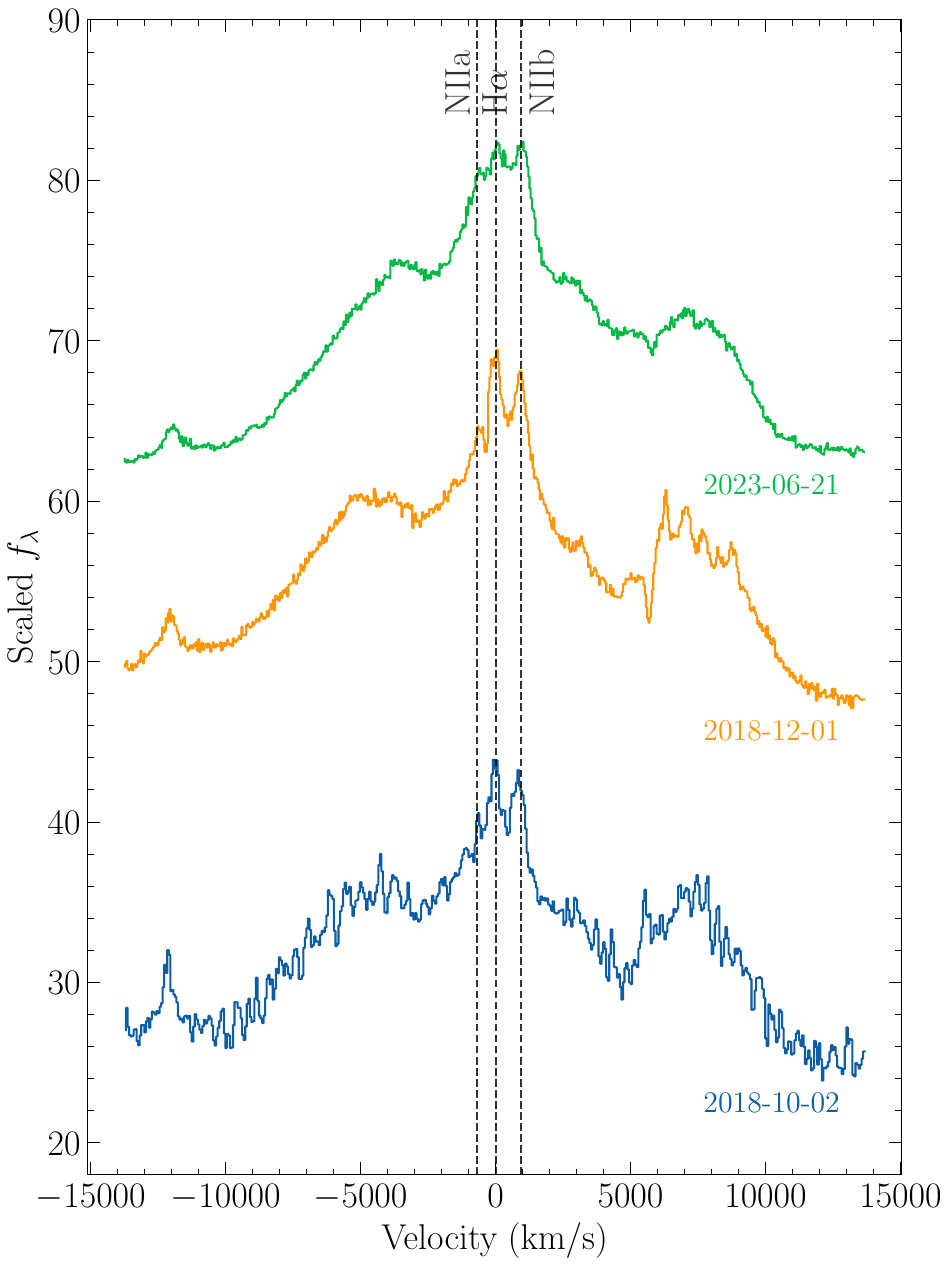}{0.34\textwidth}{}\fig{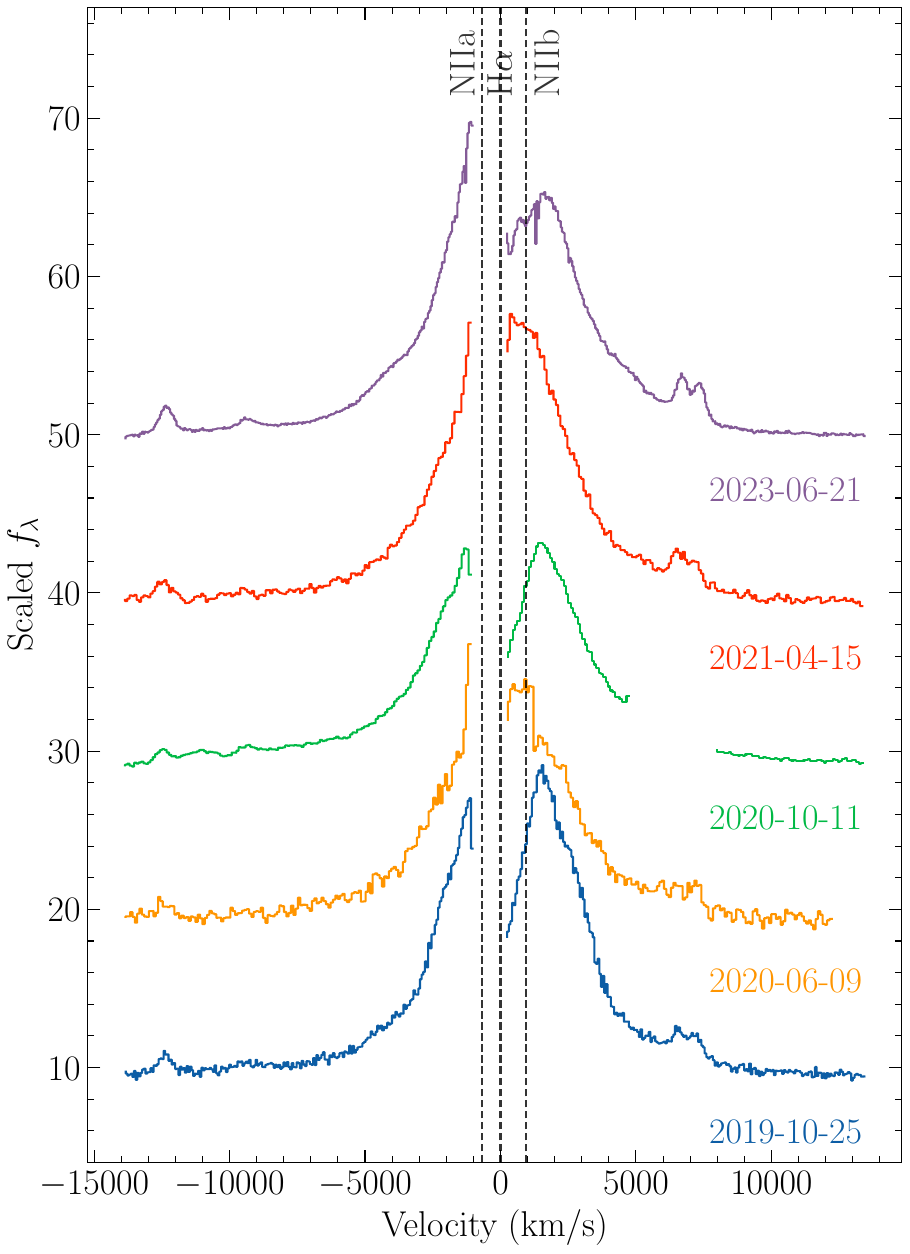}{0.32\textwidth}{}\fig{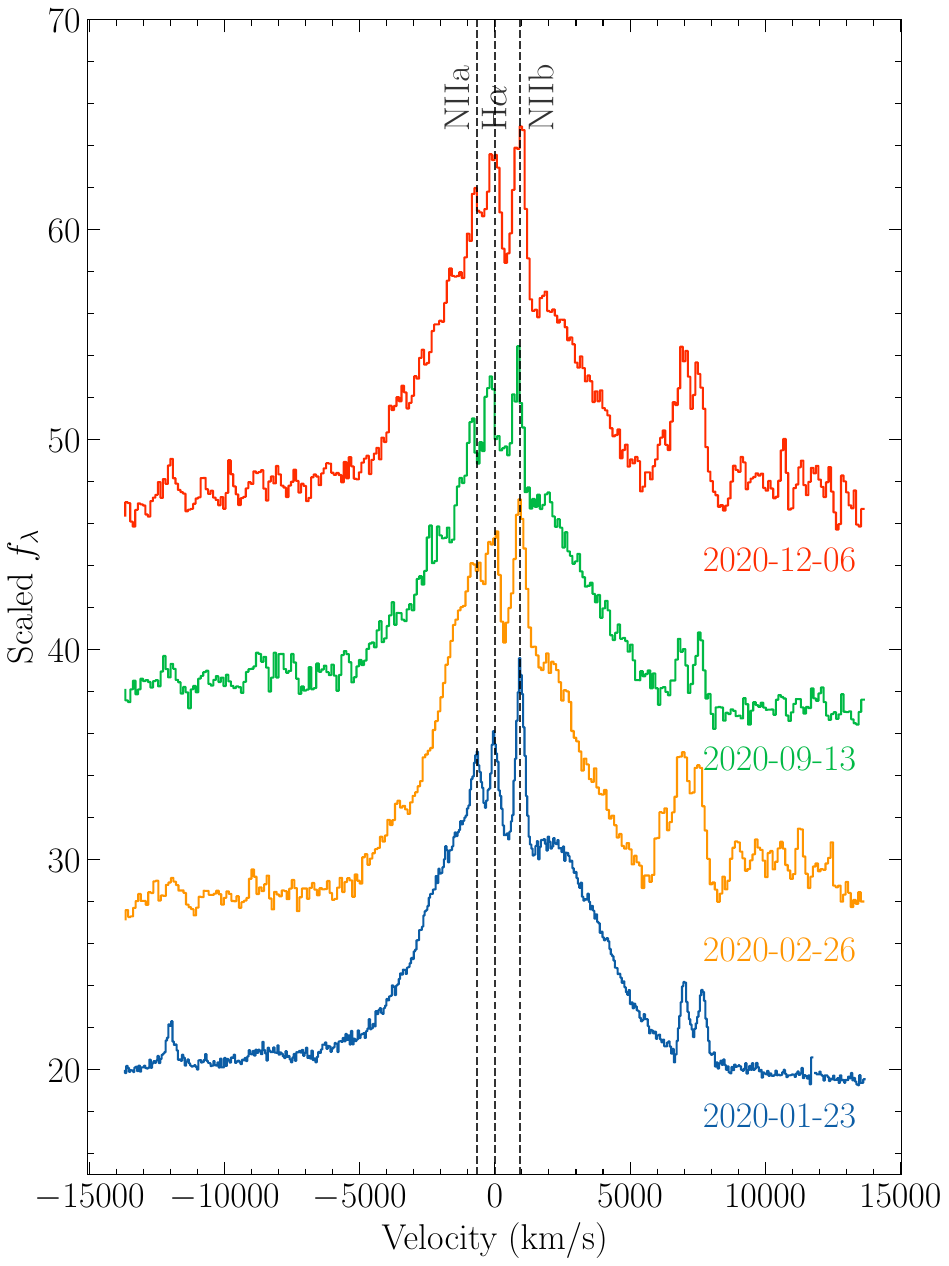}{0.33\textwidth}{}}
\gridline{\fig{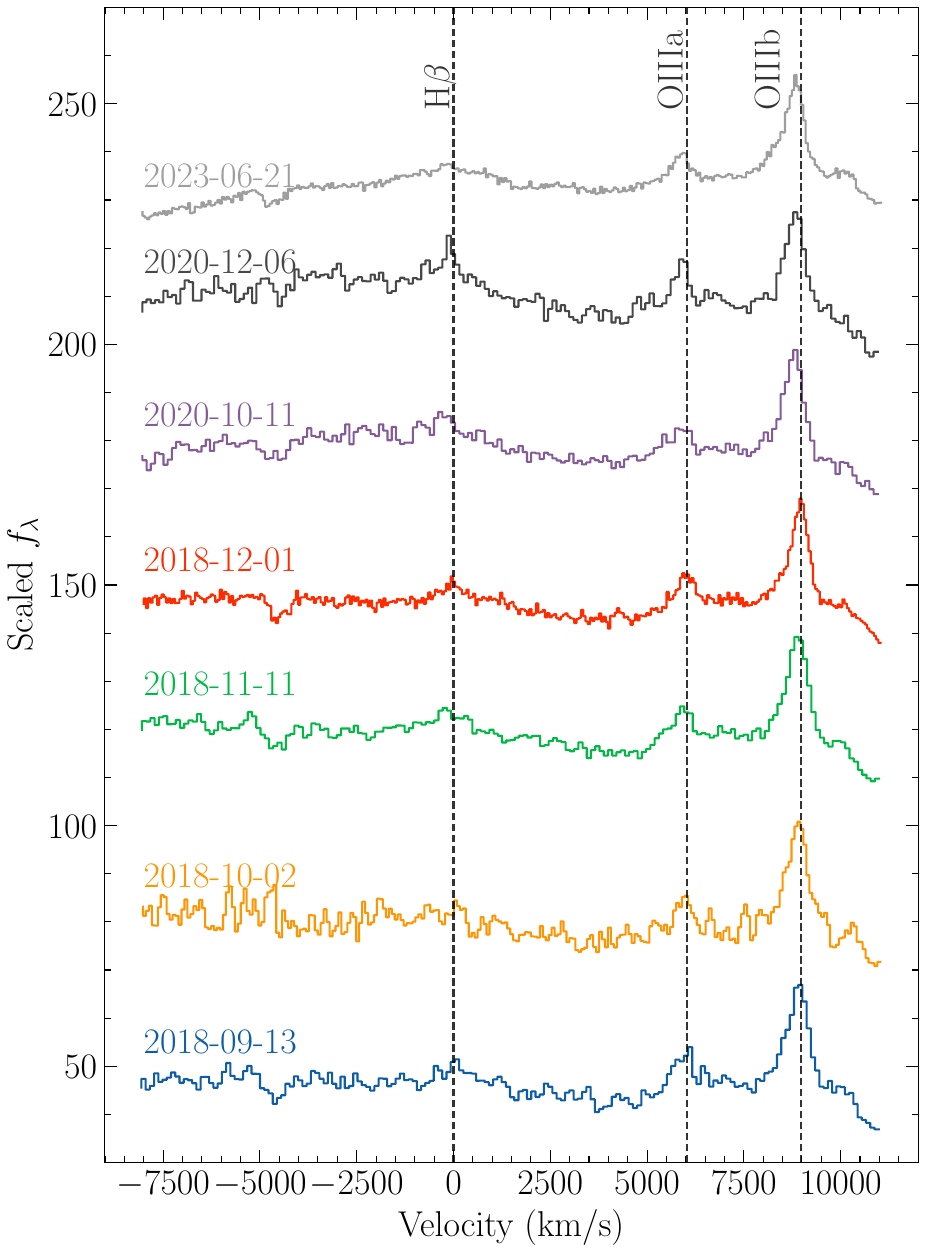}{0.33\textwidth}{ZTF18aaymybb}\fig{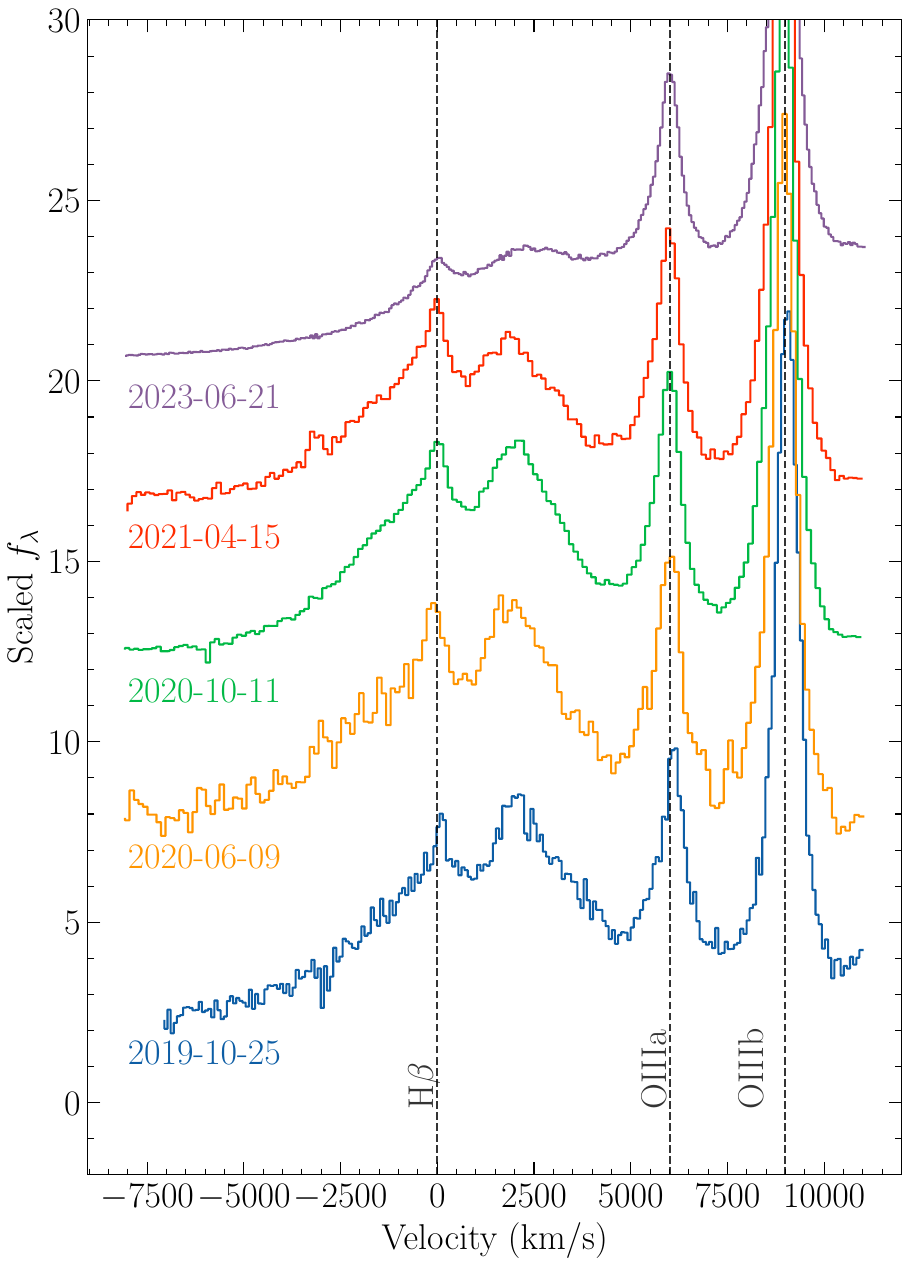}{0.33\textwidth}{ZTF18abxxohm}\fig{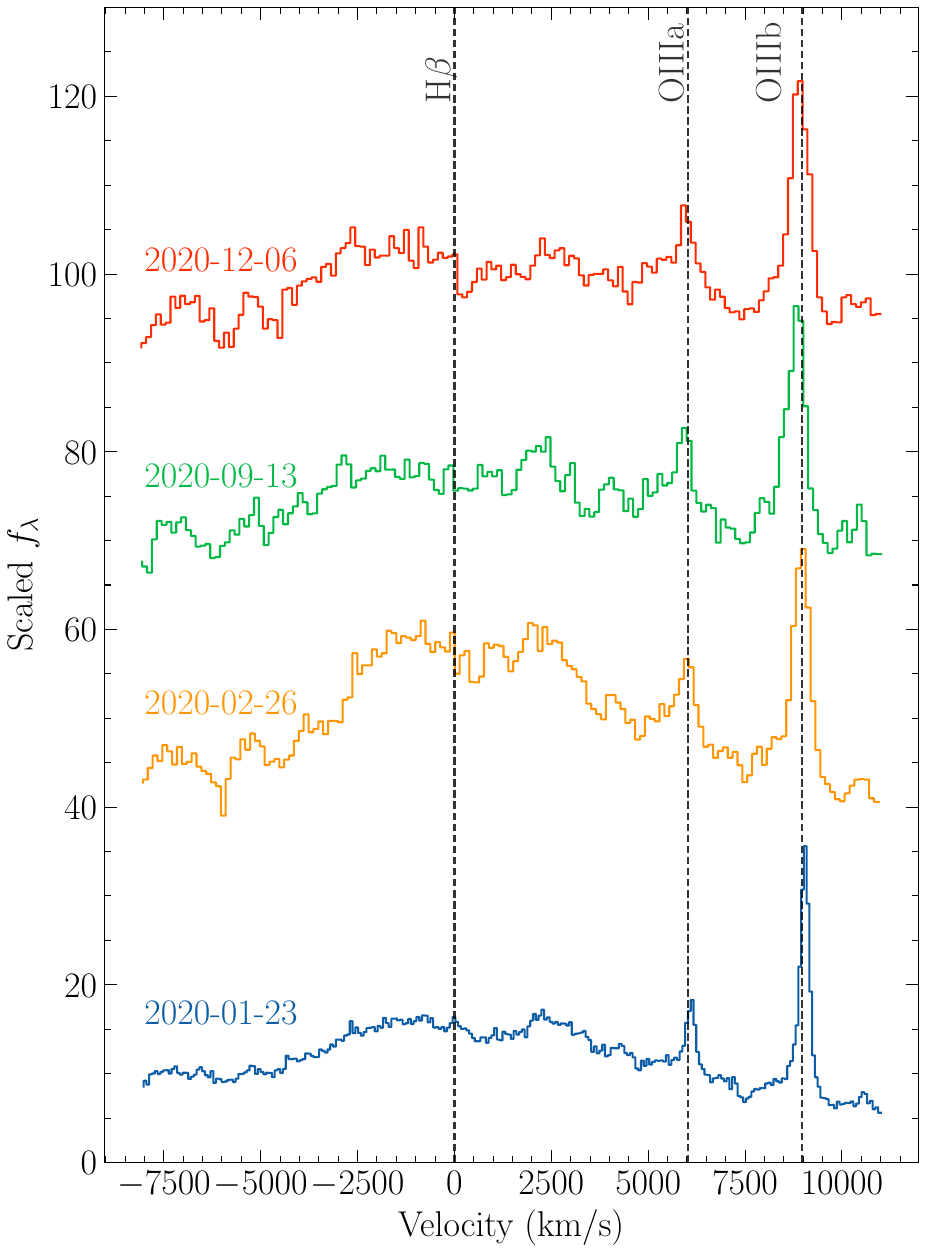}{0.33\textwidth}{ZTF19aagwzod}}
\caption{H$\alpha$ (\texttt{above}) and $H\beta$ (\texttt{below}) spectra from LDT and Keck monitoring of 3 new DPEs. The H$\alpha$ spectrum of ZTF18abxxohm has masked regions due to the presence of a strong telluric absorption line.}
\label{fig:monitor}
\end{figure*}

We took 4-7 follow-up spectra of ZTF18aaymybb, ZTF18abxxohm and ZTF19aagwzod with LDT DeVeny and Keck LRIS during 2018 to 2023 to search for changes to the profiles on the timescales of months to years. These spectra are shown in Figure \ref{fig:monitor}. ZTF19aagwzod and ZTF18aaymybb did not show significant changes in the flux of the blue and red shoulders, although ZTF18aaymybb showed changes in the shape and peak velocity of the blue shoulder between 2018 and 2023. We attribute the small, time-varying, spiky structures on the red shoulder of the ZTF18aaymybb to imperfect removal of the telluric H$_2$O absorption bands in the wavelength range 8100-8300$\AA$. 

ZTF18abxxohm exhibited a gradual decrease in the flux ratio between the red broad line and the central broad line over the course of 4 years. This is most obvious in the H$\beta$ profile evolution in the bottom row of Figure \ref{fig:longmonitor}. The peak velocity of the red broad line also varied by a few hundred km s$^{-1}$ over the course of the 4 years. 

\section{Discussion}
\subsection{Comparing the DPE and control AGN populations}
The disk parameter distributions that we find for our optically variable DPE population are similar to  those found in a previous spectroscopically-selected sample of 116 DPEs in SDSS \citep{Strateva2003Double-peakedNuclei}. While the inner radii of our DPE sample extend to values as high as $\xi_1\sim 1800$, most DPEs have inner radii within the $50<\xi_1<800$ range found in the spectroscopically selected sample. While we do find DPEs at high inclinations in the range $30<i<50$\textdegree, the majority of the population have inclinations in the range $15<i<30$, which is in agreement with both the disk modeling results of \citet{Strateva2003Double-peakedNuclei} and studies of disk inclination angles from Fe K$\alpha$ lines \citep{Nandra1997OnNuclei}. This is consistent with the overall picture that DPEs are usually only detectable at inclinations $i>15$\textdegree when the shoulders have sufficient separation, but are increasingly obscured by the dusty torus at inclinations $i>30$\textdegree \citep{Storchi-Bergmann2017Double-PeakedNuclei}.

\begin{figure*}[h]
\gridline{\fig{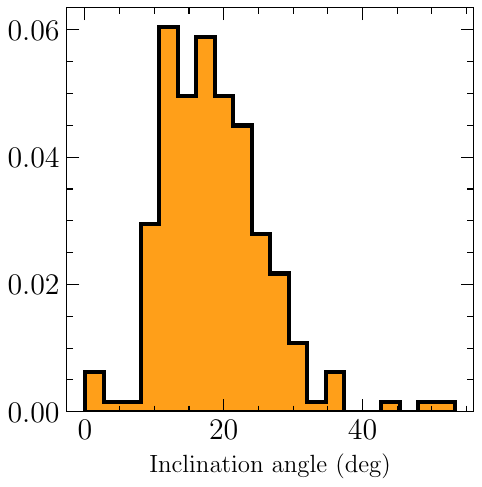}{0.28\textwidth}{} 
 \fig{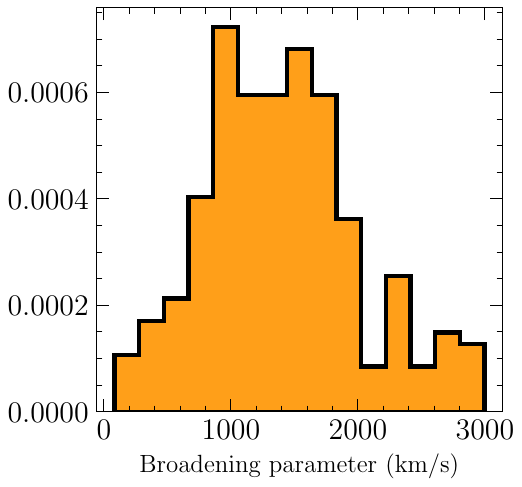}{0.3\textwidth}{} \fig{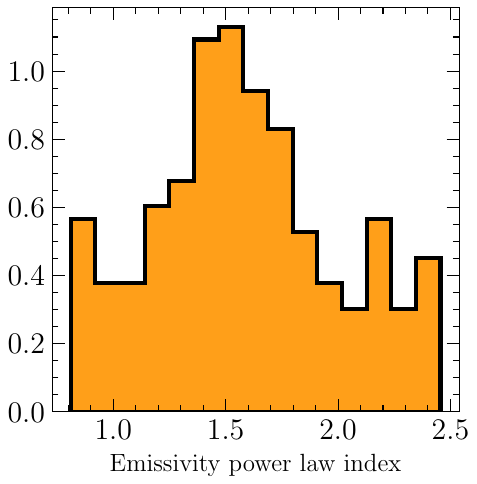}{0.28\textwidth}{}}
 \gridline{\fig{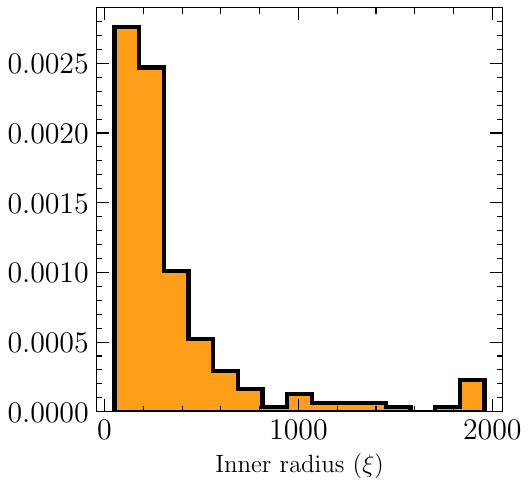}{0.3\textwidth}{} 
 \fig{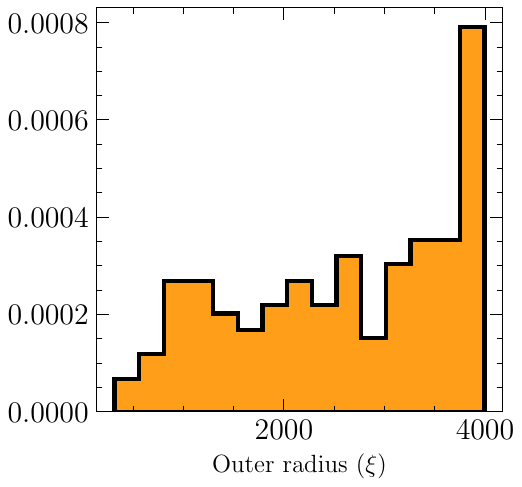}{0.3\textwidth}{} \fig{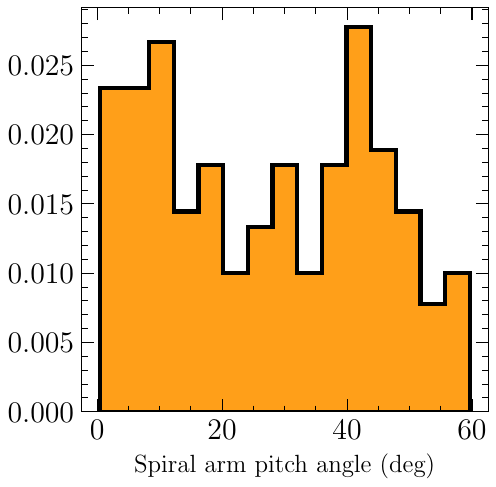}{0.3\textwidth}{}}
  \gridline{\fig{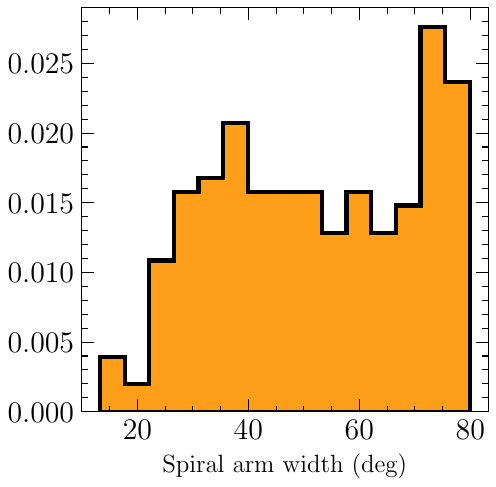}{0.3\textwidth}{} 
 \fig{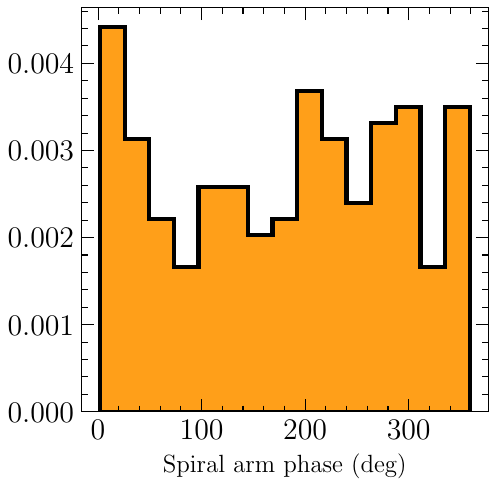}{0.3\textwidth}{} \fig{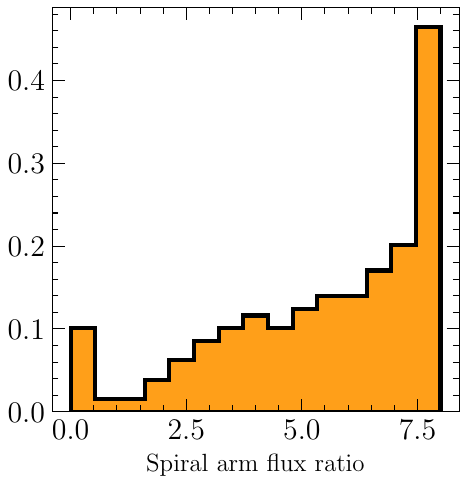}{0.3\textwidth}{}}
\caption{Distributions of the best-fit circular disk parameters for the sample of 250 DPEs. For spiral arm parameters pitch angle, width and phase, we only plot the best-fit parameters for spectra where the spiral arm contrast amplitude was greater than 1.}
\label{fig:diskDPEvsAGN}
\end{figure*}

We note that the best-fit parameters for the DPE sample show a strong preference for high spiral arm amplitudes \ref{fig:diskDPEvsAGN}. The spiral arm model is flexible enough that it can represent a very wide variety of brightness distributions \citep[e.g.][]{Schimoia2012ShortNGC1097} and may therefore simply be a useful parameterization of different phenomena causing time-varying asymmetric structures in the disk such as irradiation induced warps. For DPEs like ZTF18accwjao, which has such a large peak flux ratio between the blue and red peaks such that it appears to have a single velocity-offset broad-line, a spiral arm contrast ratio at the limit of $A_s=8$ was insufficient to fully describe the observed double-peaked profile. However, when a greater value of $A_s$ was allowed, the disk model was able to account for the large flux ratio. We take this to imply that the disk of ZTF18accwjao has an internal structure which is atypical for most DPEs, and that the circular model therefore needs greater degrees of freedom to fully model the observed profile.

Our finding that ZTF DPEs are twice as likely to be radio emitters based on VLASS survey results is consistent with the FIRST detection rates of the \citet{Strateva2003Double-peakedNuclei} DPE population. However, our finding that DPEs have similar radio detection rates to the control AGN sample at longer wavelengths in the RACS survey may point to a steeper and harder radio spectrum in DPEs arising from their larger than average inclination. We note that the VLASS catalog required the presence of an associated infrared detection in \textit{WISE} data for the radio detection to appear in the catalog, while the RACS survey did not have this requirement. The longer interferometric baselines of RACS also meant that AGN with extended jets were more likely to appear in the catalog. The finding that the radio-detected control AGN were $5\times$ more likely than radio-detected DPEs to have variable radio fluxes in VLASS on 1-3 year timescales may arise due to the expected smaller disk inclination angles of the control AGN. The presence of radio jets in a fraction of DPEs provides further opportunity to relate jet and spectroscopically-fitted disk inclinations in subset of the radio-loud population \citep[e.g.][]{Gabanyi20214C18.47:Infrared}.

Our determination that DPEs have similar distributions in variability amplitude and PSD spectral index compared to other broad-line AGN suggests that the presence of a visible accretion disk is not associated with a significant change in the level of optical variability. This is once again consistent with the idea that the primary difference between DPEs and `normal' broad-line AGN is the viewing angle rather than a difference in the accretion state of the AGN. We do, however, find that the DPE population has a statistically significant difference in the location of the low frequency turnover in the power spectrum, with DPEs having their turnovers appear at a smaller median frequency of 0.8 yr$^{-1}$ in comparison to 1.1 yr$^{-1}$ for normal broad-line AGN. This finding is comparable to the results from previous analysis of 8 year light curves from the Catalina Sky Survey Data Release 2 and SDSS Stripe 82 for DPEs and control AGN respectively, where it was found that that DPEs have characteristic timescales $\sim2.7\times$ longer than other broad-line AGN \citep{Zhang2017PropertiesLines}. We now reach similar conclusions with a uniform sample of light curves across the DPE and control populations from a single time-domain survey.

Previous analysis of 67 AGN with $\lesssim20$ year baseline light curves found a strong positive correlation between AGN mass and characteristic timescale of $\tau = 107^{+11}_{-12} \text{days} \left( \frac{M_{\text{BH}}}{10^8 M_{\odot}} \right) ^{0.38^{+0.05}_{-0.04}}$  over a BH mass range of $10^4-10^{10} M_{\odot}$ \citep{Burke2021ADisks}. We therefore considered the possibility that the shorter break frequencies (longer characteristic timescales) of DPE light curves arose because the DPEs in our sample have intrinsically larger masses than the control AGN, or because AGN are more likely to be observable as DPEs at higher masses when the turbulent broadening of the gas in the disk may be larger (see our criteria for DPE classification, which required that the turbulent broadening parameter be $>600$\, km\,s$^{-1}$; Figure \ref{fig:disksplit}). To check for a mass--break frequency correlation in our ZTF power spectra, we searched for a correlation between the best-fit break frequency and the virial masses measured from broad line widths in \citet{Ho2015ASPECTRA} for the control AGN sample. We undertook this check with the control AGN sample because the the black hole mass scaling relations have been calibrated for quasars with these types of line profiles. We found no obvious correlation between break frequency and mass for the 312 AGN in the control AGN sample with available virial masses and with break frequencies $>1.8$ yr$^{-1}$ where they can be reliably measured with the 5.5 year ZTF baselines. We confirmed the lack of mass--break frequency correlation by calculating the Spearman correlation coefficient and obtaining a p-value of 0.33 for the null hypothesis that they are uncorrelated. We note that the correlation between mass and break frequency in AGN spectra was also not reproduced in a previous analysis of $\sim3$ year baseline AGN light curves from the VLT Survey Telescope \citep{DeCicco2022AAGN}. We are therefore unable to use the masses of the populations to account for the break frequency differences between the two samples with the data at hand. 

\subsection{Comparison of disk-emitting AGN with previously reported tidal disruption events and changing-state AGN with double-peaked profiles}

The distribution of disk parameters we have found for variable AGN makes a useful point of comparison to accretion disk parameters derived from modeling of broad-line profiles in transients such as tidal disruption events. Fitting of an elliptical accretion disk model to the X-ray faint TDE AT2018hyz yielded parameters $q\sim 2.0$, $i\sim52-68$, $\sigma \sim 370 - 640$, $\xi_1 \sim 1200-1800$, and $\xi_2\sim 2600-3100$ which are typical of the ZTF DPE population except for the large inner radius, large inclination angle, and small turbulent broadening parameter \citep{Hung2020Double-peakedEvent}. The atypical disk parameters for AT2018hyz may have arisen from the use of an elliptical disk model over a circular disk model, as the elliptical disk model has alternative ways to broaden the profile and account for asymmetries between peaks. The spiral arm circular disk model fitted to spectra of the repeating TDE ASASSN14ko derived disk parameters of $i=12$, $\sigma=800$,$A_s=2.0$, $p=-55$ (pitch angle), $\xi_1=80$, $\xi_2 = 1600$ which are all typical values amongst ZTF AGN \citep{Tucker2021AnAGN}. 

The broad-line evolution observed in the 12 DPEs with spectroscopic follow-up is markedly different to the evolution in both the TDE AT2018hyz \citep{Hung2020Double-peakedEvent,Holoien2019PS18kh:Disk} and the `switching-on' AGN AT2017bcc \citep{Ridley2023Time-varyingAT2017bcc}, which both showed changes in the `boxiness', width and brightness of the double-peaked profile over time, as opposed to changes to the relative strength of the blue and red peaks. This suggests that the physical processes driving accretion disk evolution in TDEs and AGN at the onset of an accretion episode are markedly different to those producing the gradual evolution observed in stable AGN disks.

\subsection{Discussion of notable objects}
\subsubsection{Inspiraling SMBH binary candidate ZTF18aaripgg}
Our population of 250 optically variable DPEs, many of which have evolving flux ratios between the red and blue shoulders of the double-peaked profile over 10-20 year timescales, provides additional context for the original hypothesis that SDSSJ1430+2303 (ZTF18aarippg) is an inspiraling SMBH binary. The newly updated light curve of ZTF18aarippg is notable for its continued possibly quasi-periodic variability, although the original orbital model fit by \citet{Jiang2022Tick-Tock:Binary} might not be able to account for the increase in peak--peak time difference in the most recent 2 cycles. 

We expect that the apparently periodic and ringing-down signals arising in some DPE light curves are the ‘phantom’ periodicities arising naturally from correlated noise in the AGN light curves. Simulations of light curves arising from DRW power spectra with $\tau=200$ days and slopes of $\gamma=2$ in 9-yr CRTS datasets by \citet{Vaughan2016FalseSurveys} found that $\sim 1-2$ per 1000 light curves showed periodic behaviour well-fit by a sinusoidal model. Furthermore, they found that the fraction with false periods increased to $\sim 1$ in 200 when the spectral index of the PSD was increased from 2 to 3. We note that large spectral indices of up to 4 were found in a fraction of ZTF AGN power spectra, and ZTF18aarippg has a relatively large power spectral index of 3.79 relative to the overall DPE and AGN distributions (Figure \ref{fig:psd_hists}).  Given the expected rates of phantom periodicities and evolving double-peaked broad lines that we find, we expect that SDSSJ1430+2303 is likely to be a single disk-emitting AGN, consistent with arguments in other follow-up studies \citep{Dotti2022OpticalCandidate}. This conclusion is also consistent with recent theoretical predictions on the spectroscopic and variability signatures of SMBH binaries which paint a pessimistic picture for the existence of kinematically observable SMBH binaries \citep{Kelley2020ConsiderationsAGN,Kelley2019MassiveAGN}.

\subsubsection{ZTF18aalslhk: A possibly quasi-periodic DPE with an evolving disk profile}
While ZTF18aalslhk appears to have quasi-periodic variability over 4-5 cycles in its optical light curve, its power spectrum properties are fairly typical for the observed DPE and AGN distributions, with the light curve having a larger than median variability amplitude but average values of the power law index and break frequency (Figure \ref{fig:psd_hists}). Its disk profile shows major structural changes likely caused by the precession of a spiral arm or hotspot. Detailed spectroscopic monitoring of this object over the next few years could test for a relationship between spiral arm phase and optical flux variations. 

\subsubsection{ZTF18aarywbt: Asymmetric radio jets and an evolving disk profile}
ZTF18aarywbt, with fitted disk inclination angle 19\textdegree, has a radio point source at the AGN location visible in 20cm VLASS images, and asymmetric, multi-lobed radio jets spanning $\sim1^{\prime}$ ($\sim10$ kpc) which are visible in both VLASS and 34cm RACS images. It is also X-ray bright \citep[e.g.][]{Ricci2017BATCatalog}. This source shows a dramatic change in the double-peaked profile between archival SDSS spectra from 2005 and follow-up LDT spectra in 2021, with the complete disappearance of the red shoulder and a decrease in the velocity of the blue shoulder. Future analysis could investigate whether the radio jet structures and disk evolution are indicative of disk precession.

\subsubsection{ZTF18aaymybb: A complex disk profile with disappearing small scale structure}
Previously unreported DPE ZTF18aaymybb has an H$\alpha$ profile which stands out amongst other disk-emitters because the shoulders are not as smooth and rounded as the majority of the DPE sample. The cuspiness of the shoulders in fact makes this object more comparable to Arp102B, the original archetypal DPE \citep{Chen1989Kinematic102B}. The peak velocity of the blue shoulder decreased by $\sim 500$ km/s between 2018 and 2023. We attribute the time-varying, narrow ($\sim50$ km/s) peaks on the red shoulder of the disk profile to imperfect removal of telluric H$_2$O absorption features, rather than real features which may arise due to shocks or local disk motions \citep{Gezari2007LongTermNuclei,Lewis2010Long-termLines}, because they do not appear in the H$\beta$ profile as well. The ZTF light curve power spectrum of ZTF18aaymybb is fairly typical for the DPE/AGN population (Figure \ref{fig:powerspec}).

\subsubsection{ZTF18abxxohm: An unusual and evolving velocity-offset broad line}

Newly identified DPE ZTF18abxxohm does not have a typical double-peaked profile which is well described by a circular disk model, instead showing one broad line at the rest velocity and a second broad line, with no associated narrow lines, at $\sim2000$ km/s red of the rest velocity. The red shoulder exhibited a gradual decrease in the flux over the course of 4 years (Figure \ref{fig:longmonitor}). The variation in the peak velocity of the red shoulder by a few hundred km s$^{-1}$ may be attributable to radial velocity jitter arising from fluctuations in the continuum which illuminates the broad line gas \citep{Barth2015THECurves, Guo2019ConstrainingTests, Doan2020AnLines}. This object has a fairly typical AGN light curve for this population, with a g-band power spectral index of 3.13. ZTF18abxxohm may be worthy of further follow-up as an SMBH binary candidate or an AGN with a high velocity outflow. 

\subsubsection{ZTF19aagwzod: A previously reported CLAGN candidate with a stable disk profile}

ZTF19aagwzod (LEDA 1154204) was originally reported as a candidate CLAGN due to its dramatic increase in magnitude from g=19.6 mag to g=17.9 mag over 34 days observed in December 2019 \citep{Frederick2020AX-rays}. The initial discovery report also noted in 2019 that it was UV and X-ray bright, with an X-ray power law spectral index of $1.8 \pm 0.1$ consistent with an AGN \citep{Frederick2020AX-rays}. The four follow-up spectra taken from 2020-01-23 to 2020-12-06 show a stable disk profile with very little change over the course of the year following the flare (Figure \ref{fig:longmonitor}). The combination of optical BPT line ratios indicating that AGN activity has persisted for millennia prior to this event and lack of X-ray spectral evolution led \citet{Saha2023MultiwavelengthSeyfert} to conclude that the change of classification from Seyfert 1.9 to Seyfert 1 associated with the optical flare was due to a temporary instability in the accretion flow. The lack of evolution in the disk profile suggests that the disk structure stabilized quickly after the optical flare. This makes an interesting comparison to AT2017bcc, which showed substantial evolution in the disk profile shape following an optical flare in a previously quiescent galaxy with no evidence for previous AGN activity in the BPT emission line classifications \citep{Ridley2023Time-varyingAT2017bcc}. 

\section{Summary and conclusions}

We have presented a population of 250 optically variable AGN with double-peaked or velocity-offset H$\alpha$ and H$\beta$ broad-line profiles consistent with emission from a circular accretion disk. We found that 19.2\% of broad-line AGN in ZTF with $>1.5$ magnitude optical variability are DPEs. We have modeled the H$\alpha$ broad line regions as circular disks for both the DPE sample and control AGN population, and provide a catalog of disk parameters for the 250 DPEs. We have presented the distributions of best-fit disk parameters for the DPE population.

We generated power spectra of the DPE and control AGN light curves and fit them with a power law model with a low frequency turnover and a high frequency intrinsic white noise component. We have provided a catalog of the power spectrum parameters derived from the ZTF light curves. We found that DPEs and other broad-line AGN have similar distributions in PSD amplitude and power law spectral index, but that DPEs tend to have a turnover in their power spectra at lower frequencies (and therefore longer characteristic timescales) than other broad-line AGN.

We have shown that DPEs have much higher radio detection fractions at 20cm wavelengths than the control AGN sample, but that this is not the case at 34cm wavelengths. We presented radio imaging of three DPEs with notable jet structures associated with accretion disks at inclinations of $\sim14-35$\textdegree.

Spectroscopic follow-up of 12 DPEs indicated that $\sim50$\% show significant changes in the relative strengths of the blue and red shoulders over $\sim15$ year timescales (Figure \ref{fig:longmonitor}). There are also many examples of DPE and other AGN light curves that appear to show quasi-regular fluctuations in their 4-year optical light curves. We therefore conclude that previously reported SMBH binary candidate ZTF18aarippg does not have unusual broad-line evolution or light curve properties compared to the larger optically variable DPE population. The population statistics presented in this paper could be used to inform future calculation of false positive rates for selection of SMBH binary candidates using optical light curves and time-resolved spectroscopy. We have also shown that the broad-line evolution typical of DPEs in the ZTF AGN sample is different to the evolution observed in the TDE AT2018hyz and the `switching-on' AGN AT2017bcc, indicating that different physical processes are driving the changes to accretion disk structure in the newly formed accretion disks in these transient events.

Our sample of DPEs exhibits only minor differences in optical variability behavior compared to the remaining broad-line AGN in our ZTF sample. The results of our spectroscopic and light curve analysis are consistent with the interpretation that DPEs do not have major physical differences to other broad-line AGN, and their differing spectroscopic and radio properties may merely arise from selection effects such as preferences for intermediate disk viewing angles. 

As part of this paper, we have made available the regularly-sampled 5 year optical ZTF light curves and the corresponding mid-IR \textit{WISE} light curves showing delayed dust echos. We have also produced catalogs of spectroscopically-derived accretion disk geometry parameters, radio fluxes, and optical power spectrum properties for 250 DPEs. We have presented spectra, light curves and radio jet imaging of unusual DPEs which may be worthy of further investigation. As time-domain surveys like ZTF -- and, in the near future, the Legacy Survey of Space and Time at the Vera 
C. Rubin Observatory \citep{Ivezic2019LSST:Products} -- continue to discover evolving DPEs, changing-state AGN, and tidal disruption events with disk-like profiles, we hope that this population study provides useful context as we work to understand the ways SMBH accretion disks form and evolve in various scenarios.

\section{Acknowledgements}

We would like to thank the anonymous referee for their helpful and constructive feedback. Based on observations obtained with the Samuel Oschin Telescope 48-inch and the 60-inch Telescope at the Palomar Observatory as part of the Zwicky Transient Facility project. ZTF is supported by the National Science Foundation under Grant No. AST-2034437 and a collaboration including Caltech, IPAC, the Weizmann Institute for Science, the Oskar Klein Center at Stockholm University, the University of Maryland, Deutsches Elektronen-Synchrotron and Humboldt University, the TANGO Consortium of Taiwan, the University of Wisconsin at Milwaukee, Trinity College Dublin, Lawrence Livermore National Laboratories, and IN2P3, France. Operations are conducted by COO, IPAC, and UW. The ZTF forced-photometry service was funded under the Heising-Simons Foundation grant 12540303 (PI: Graham). This work was supported by the GROWTH project \citep{Kasliwal_2019} funded by the National Science Foundation under Grant No 1545949.

This work was supported by a NASA Keck PI Data Award, administered by the NASA Exoplanet Science Institute. Data presented herein were obtained at the W. M. Keck Observatory from telescope time allocated to the National Aeronautics and Space Administration through the agency's scientific partnership with the California Institute of Technology and the University of California. The Observatory was made possible by the generous financial support of the W. M. Keck Foundation. This research has made use of the Keck Observatory Archive (KOA), which is operated by the W. M. Keck Observatory and the NASA Exoplanet Science Institute (NExScI), under contract with the National Aeronautics and Space Administration.

This research has made use of the CIRADA cutout service at URL cutouts.cirada.ca, operated by the Canadian Initiative for Radio Astronomy Data Analysis (CIRADA). CIRADA is funded by a grant from the Canada Foundation for Innovation 2017 Innovation Fund (Project 35999), as well as by the Provinces of Ontario, British Columbia, Alberta, Manitoba and Quebec, in collaboration with the National Research Council of Canada, the US National Radio Astronomy Observatory and Australia’s Commonwealth Scientific and Industrial Research Organisation.

This publication also makes use of data products from NEOWISE, which is a project of the Jet Propulsion Laboratory/California Institute of Technology, funded by the Planetary Science Division of the National Aeronautics and Space Administration.

This research has made use of the NASA/IPAC Infrared Science Archive, which is funded by the National Aeronautics and Space Administration and operated by the California Institute of Technology.

This research used resources of the National Energy Research Scientific Computing Center, a DOE Office of Science User Facility supported by the Office of Science of the U.S. Department of Energy under Contract No. DE-AC02-05CH11231. P.E.N. acknowledges support from the DOE under grant DE-AC02-05CH11231, Analytical Modeling for Extreme-Scale Computing Environments.

Work at NRL is supported by NASA.

\software{
Ampel \citep{Nordin2019TransientCurves},
Astropy \citep{TheAstropyCollaboration2022ThePackage}, 
GROWTH Marshal \citep{Kasliwal_2019},
pPXF \citep{Cappellari2003, Cappellari2017ImprovingFunctions}
}

\bibliography{main.bib}{}
\bibliographystyle{aasjournal}

\section{Appendix}

\end{document}